%% file: main.tex
\documentclass[journal]{IEEEtran}
\ifCLASSINFOpdf
\else
\fi
\usepackage{amsthm}
\usepackage{amsmath}
\usepackage{color}
\usepackage{amssymb}
\usepackage[official]{eurosym}
\usepackage[nodisplayskipstretch]{setspace}
\usepackage{xspace}
\usepackage{array,multirow,colortbl,bigdelim,hhline}
 
\usepackage{booktabs}
\usepackage{balance}

\usepackage{arydshln}

\usepackage{makecell}

\usepackage[font=footnotesize]{caption}
\usepackage[font=footnotesize]{subcaption}

\usepackage{nohyperref}
\usepackage[hyphens]{url}
\usepackage{mathtools}
\usepackage{nicefrac}
\usepackage[table,dvipsnames]{xcolor}

\let\originalleft\left
\let\originalright\right
\renewcommand{\left}{\mathopen{}\mathclose\bgroup\originalleft}
\renewcommand{\right}{\aftergroup\egroup\originalright}

\newcolumntype{L}[1]{>{\raggedright\let\newline\\\arraybackslash\hspace{0pt}}m{#1}}
\newcolumntype{C}[1]{>{\centering\let\newline\\\arraybackslash\hspace{0pt}}m{#1}}
\newcolumntype{R}[1]{>{\raggedleft\let\newline\\\arraybackslash\hspace{0pt}}m{#1}}

\usepackage[normalem]{ulem}

\newcommand{\eg}{\textit{e.g.},\xspace}
\newcommand{\ie}{\textit{i.e.},\xspace}
\newcommand{\etal}{\textit{et al.}\xspace}
\newcommand{\comma}{\mathbin{\raisebox{0.5ex}{,}}}

\RequirePackage{tikz}
\usetikzlibrary{calc,positioning,shapes,shadows,arrows,fit,decorations.pathreplacing,backgrounds,matrix,patterns}
\usepackage{tikz-qtree}

\begin{document}
%
\title{Mobility as an Alternative \\ Communication Channel: A Survey}
%
%
%

\author{Benjamin Baron, Prom{\'{e}}th{\'{e}}e Spathis, Marcelo Dias de Amorim, Yannis Viniotis, and Mostafa H. Ammar
\thanks{Benjamin Baron, Prom{\'{e}}th{\'{e}}e Spathis, and Marcelo Dias de Amorim are with the LIP6/CNRS Computer Science  laboratory, Universit{\'{e}} Pierre et Marie Curie, Paris, France. Emails: \{bbaron,spathis,amorim\}@npa.lip6.fr. Yannis Viniotis is with North Carolina State University, NC, USA. Email: candice@ncsu.edu. Mostafa H. Ammar is with Georgia Institute of Technology, GA, USA. Email: ammar@cc.gatech.edu.}}

\markboth{Baron \MakeLowercase{\textit{et al.}}: Mobility as an Alternative Communication Channel: A Survey}%
{Baron \MakeLowercase{\textit{et al.}}: Mobility as an Alternative Communication Channel: A Survey}
%



\maketitle

\begin{abstract}

We review the research literature investigating systems in which mobile entities can carry data while they move. These entities can be either mobile by nature (\eg human beings and animals) or mobile by design (\eg trains, airplanes, and cars). The movements of such entities equipped with storage capabilities create a communication channel which can help overcome the limitations or the lack of conventional data networks. Common limitations include the mismatch between the capacity offered by these networks and the traffic demand or their limited deployment owing to environmental factors. Application scenarios include offloading traffic off legacy networks for capacity improvement, bridging connectivity gaps, or deploying ad hoc networks in challenging environments for coverage enhancement. 

\end{abstract}

%
\IEEEpeerreviewmaketitle

\input{1a-introduction.tex}
\input{1b-use-cases.tex}

\input{1c-mobile-entities.tex}

\input{3a-direct-delivery.tex}
\input{3b-indirect-async.tex}

\input{3c-indirect-sync.tex}

\input{4-discussions.tex}

\input{5-conclusion.tex}

%
%

\bibliographystyle{IEEEtran}
\bibliography{survey.bib}

%

%
%
%




\end{document}

%% file: 1a-introduction.tex
\section{Introduction}

Traditional data delivery approaches such as infrastructure-based networks are now ubiquitous. They allow transferring large amounts of data such as movies and enable real-time communications via voice or video. While these approaches are well provisioned to handle the communication needs of a large proportion of the world's population, they fall short in two key domains.
    
Firstly, the traditional approaches are inadequate for handling transfers of massive amounts of data. Indeed, the growth of data traffic might bring the Internet to a \textit{capacity crunch} in the near future~\cite{hecht2016bandwidth,index2016forecast}. With the enhancement of access networks and the demand increase in data traffic, content delivery network providers such as Akamai have reported that the bottleneck is no longer at the origin or the destination of the transfers, but could also be at the core of the network, including peering points between Internet Service Providers and within provider networks~\cite{nygren2010akamai}. Furthermore, as reported by Hecht, geo-distributed services are one of the biggest drivers of demand for bandwidth, as they require transfers of massive amounts of data for synchronization and maintenance~\cite{hecht2016bandwidth}. While network operators can address this issue by extending the capacity of their networks (\eg with high-throughput dedicated lines or additional peering connections), these infrastructure upgrades do not generally scale and have a poor cost-efficiency. In this context, finding alternative ways to transfer such data becomes paramount for service providers that operate large-scale networks and scientific communities that continuously exchange massive amounts of data among several research centers (\eg astronomy, remote sensing, and physics).

    
Secondly, the traditional approaches can have limited coverage, a fact commonly known as \textit{digital divide}~\cite{norris2001digital}. This limitation is mainly due to the fact that deploying networking infrastructure is costly and the return on investment could be very low in areas with low densities of population such as rural areas. Deploying infrastructure could also be impossible due to political reasons, for instance in a conflict area or on battlefields, or due to environmental reasons when the area does not lend itself to building infrastructure, for instance with disaster relief, dense remote forests and deep seas. As such, organizations that wish to set up a communication network in these challenging environment must turn to alternative data transportation systems.

In summary, the real world problem is posed by the \textit{lack of coverage} of existing communication infrastructure, as well as the limited capacity of the infrastructure to support the transport of \textit{massive} amounts of data inexpensively.

A technique that could be used to solve the problem exploits the mobility of entities naturally present in these areas to transfer data. The use of mobile entities becomes natural to handle large-scale data transfers, as they can provide the same level of service as legacy infrastructure-based networks in terms of security and throughput, while not consuming the scarce bandwidth of legacy networks. Depending on the entities used, the data transfers that take place have different levels of reliability, as the movements of these entities can be random, completely or partially predictable, or fully controlled.



\begin{figure}
    \centering
    \includegraphics[width=\columnwidth]{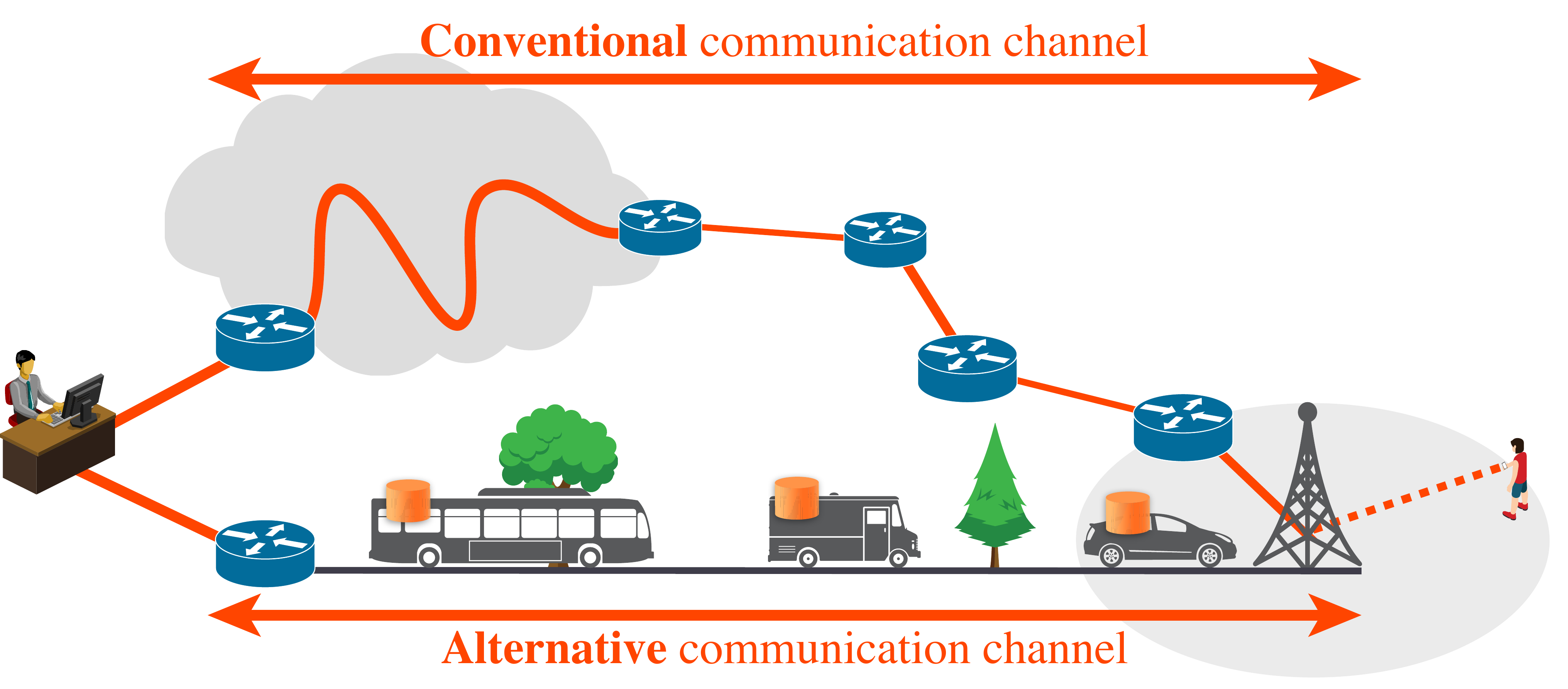}
    \caption{Mobility as as alternative communication channel: The movements of various mobile entities together with a storage device can be used to transfer data and replace or assist a conventional data network.}
    \label{fig:alternative-channel}
\end{figure}

In this survey, we are interested in the alternative data transmission methods enabled by the piggybacking of data on storage devices within mobile entities. As depicted in Figure~\ref{fig:alternative-channel}, these methods, informally referred to as Sneakernets~\cite{gray2003conversation}, exploit the movements of mobile entities to transfer data in replacement to or to assist conventional transmissions over computer networks. Typical use-case scenarios and applications consider specific types of data that can tolerate delay in their delivery. Those scenarios involve a wide range of entities that are mobile either by nature, such as humans or animals, or by conception, such as motor vehicles or robots. 

Depending on the entities involved and the services exploiting their movements, the expected benefits of these alternative transmission methods depend on various parameters such as the total number of mobile entities and the size of the memory they carry. 
Implementing such communication channels using mobile entities presents additional technical and scientific challenges. A common challenge addressed by works we review is to maximize data transfer performance by alleviating the limited control or the lack of knowledge regarding the mobility of the entities, especially when these entities move for purposes other than the intended data exchange~\cite{balasubramanian2007dtn,baron2017centrally}. While the approaches presented in the survey are relatively new and not mature, we aim to analyze the technical challenges they address by answering the following questions of interest:
\begin{itemize}
    \item \textit{When designing an alternative communication channel, what data and which entities should one consider, given the channel requirements?}
    
    \item \textit{When administrating a channel formed by moving entities, is it possible to efficiently configure and monitor its performance?}
    
    \item \textit{What are the expected benefits of deploying such alternative channels as compared to an infrastructure-based network?}
\end{itemize}

To answer these questions and related challenges, the different approaches develop and use various performance metrics we will detail in Section~\ref{sec:perf-metrics}. These metrics include the data delivery delay, data delivery ratio, and the protocol overhead. However, not all metrics are used in the papers we survey, because of the various trade-offs made by the authors that make the metrics not relevant with respect to their strategies.

\begin{figure}
    \centering
    \begin{subfigure}[b]{\columnwidth}
        \centering
        \includegraphics[width=\columnwidth,page=2]{figures/data-delivery-typesv2.pdf}
        \caption{\textbf{Direct} data delivery.}
        \label{fig:data-delivery-types-direct}
    \end{subfigure}
    
    \begin{subfigure}[b]{\columnwidth}
        \centering
        \includegraphics[width=\columnwidth,page=1]{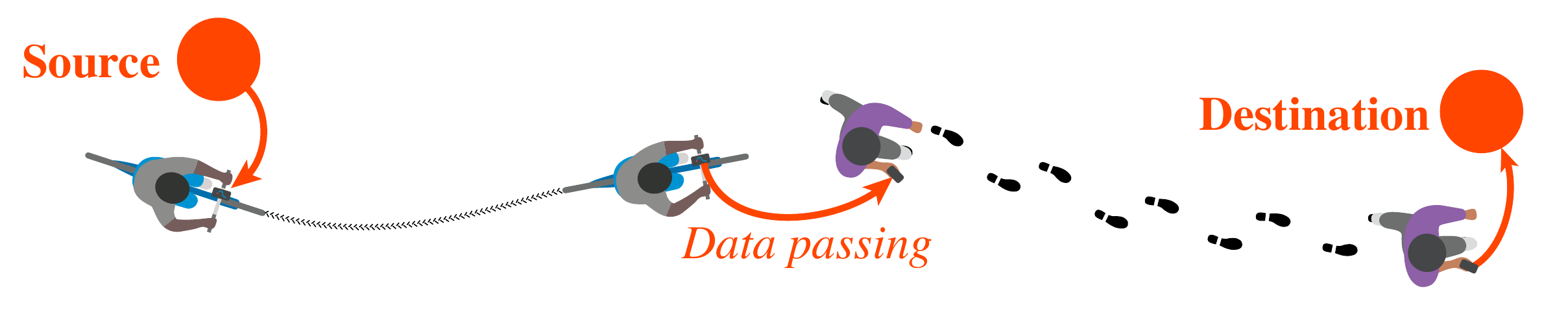}
        \caption{\textbf{Indirect} data delivery.}
        \label{fig:data-delivery-types-indirect}
    \end{subfigure}

    \caption{Representation of (a) direct and (b) indirect data delivery approaches.}
    \label{fig:data-delivery-types}
\end{figure}

Our survey consists of two parts depicted in Figure~\ref{fig:data-delivery-types}. In the first part of the survey, we review the direct data delivery approaches (Figure~\ref{fig:data-delivery-types-direct}) that exploit the mobility of \textbf{one or more independent entities to carry data directly} between a source and a destination. The main difference between these approaches lies in the degree of randomness in the movements of the entities they employ. We classify the direct data delivery approaches according to the purpose of the entity movement. We distinguish the approaches that (\textit{i}) passively exploit the entities' existing mobility, from those that (\textit{ii}) actively relying on controllable entities, and those (\textit{iii}) using a paid service such as a parcel delivery service. In the first case, the entities move for other purposes than delivering the data they carry with more or less predictable movements. In the second case, the entities follow trajectories calculated for the specific purpose of data delivery. In the third case, the entities refer to the fleet of vehicles operated by a postal or package delivery company in charge of transporting the data.

In the second part of the survey, we review the indirect data delivery approaches (Figure~\ref{fig:data-delivery-types-indirect}) that \textbf{involve a sequence of mobile entities that take turns in delivering the data}. The data takes a logical path consisting of multiple segments of trajectories, each followed by different mobile entities. Data is passed from one entity to another as a result of a process referred to as \textit{forwarding} when entities are in direct contact. We classify these approaches based on the strategy they use for composing the trajectories of the different entities. These strategies are a combination of the two following criteria: 
\begin{enumerate}
    
    \item The \textit{time} when the composition occurs. The data can either be passed synchronously at the time two entities meet or asynchronously by buffering it at specific locations.  
    
    \item The \textit{location} where the composition is performed. The location can be either pre-positioned or floating. In the pre-positioned case, the data is passed only if the entities are in contact at specific locations. In the floating case, the composition results from contacts between entities wherever they meet.  

\end{enumerate}

The main contributions of this survey are the following:
\begin{itemize}
  
  \item We propose a comprehensive classification that covers a wide variety of approaches exploiting mobile entities for delivering data in replacement of or to assist traditional data networks.
  
  \item Our classification covers the methods used to select the sequence of entities whose movements result in a vector route followed by the data and to pass the data between adjacent entities.
  
  \item We review the methods and tools developed to analyze and characterize the mobility of various entities using real-world mobility traces. 

\end{itemize}

To the best of our knowledge, this survey is the first to tackle mobility from the perspective of the communication channel resulting from the movements of entities regardless of their nature. Whereas other surveys have focused on the forwarding strategies or the impact of mobility models on the delivery success in specific networks such as Disruption-Tolerant Networks or DTNs~\cite{zhang2006routing,pereira2012delay,spyropoulos2010routing}, our survey covers a wide spectrum of network models including but not limited to DTNs or techniques such as forwarding or routing in those networks.

The structure of this paper is as follows. In Section~\ref{sec:usecase}, we present the motivations of the works we survey and use case scenarios that benefit from the communication channel resulting from the movements of entities carrying data. We classify these entities in Section~\ref{sec:entities}. In Section~\ref{sec:direct-delivery}, we give an overview of the direct data delivery strategy. We then present the three direct data delivery approaches based on the purpose of the entity movement: (\textit{i}) the passive method in Section~\ref{sec:direct-non-control}, (\textit{ii}) the active method in Section~\ref{sec:direct-control}, and (\textit{iii}) the paying method in Section~\ref{sec:contracted-mobility}. In Section~\ref{sec:indirect-delivery}, we give an overview of the indirect data delivery approach which involves a sequence of relay mobile entities. We present a classification of the approaches following the indirect approach depending on the method they introduce to pass the data between two consecutive entities as a combination of the time and location criteria. In Section~\ref{sec:indirect-async-stationary}, we present those relying on the asynchronous data passing via stationary intermediate nodes and in Section~\ref{sec:indirect-async-mobile}, those relying on mobile intermediate nodes. We then present the approaches relying on the synchronous passing method in Section~\ref{sec:indirect-sync}. In Section~\ref{sec:benj-baron}, we share the experience we acquired through the design of an offloading system involving private cars as data carriers. In Section~\ref{sec:metrics-eval-tools}, we discuss the performance metrics and tools commonly used to evaluate those data delivery methods. Finally, we conclude our paper 
with an overview of open research challenges and outlooks in Section~\ref{sec:conclusion}.

%% file: 1b-use-cases.tex
\section{Target use-case scenarios \\ and data delivery challenges}
\label{sec:usecase}

\begin{figure}
    \centering
    \begin{subfigure}[b]{\columnwidth}
        \centering
        \includegraphics[width=\columnwidth,page=2]{figures/dualChannelv2.pdf}
        \caption{\textbf{Assist} existing conventional data networks.}
        \label{fig:dual-channel-assist}
    \end{subfigure}
    
    \begin{subfigure}[b]{\columnwidth}
        \centering
        \includegraphics[width=\columnwidth,page=1]{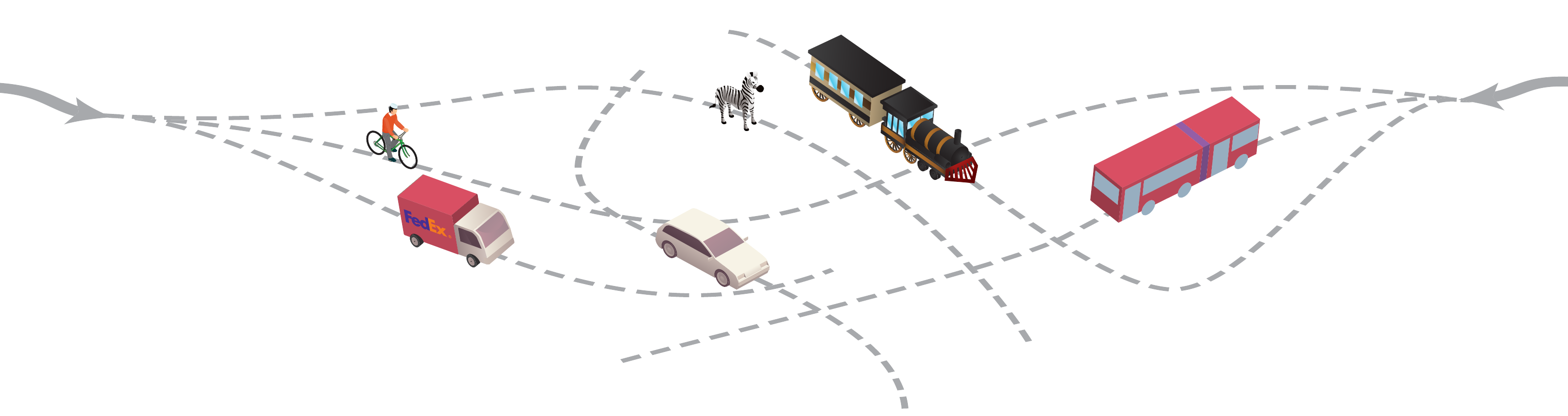}
        \caption{\textbf{Create} an alternative communication channel.}
        \label{fig:dual-channel-create}
    \end{subfigure}
    
    \caption{The movements of various entities create an alternative communication channel either used (a) to assist existing conventional data networks or (b) in replacement of such networks.}
    \label{fig:dual-channel}
\end{figure}

The work covered in this survey has been conducted with use-case scenarios that rely on the communication channel resulting from the movements of a wide range of entities carrying data. The entities are not necessarily computer devices~---~they can be arbitrary mobile entities equipped with storage capabilities. In the following, we detail two use-case scenarios depicted in Figure~\ref{fig:dual-channel} where traditional data delivery approaches fall short and require operators to rely on mobility-based approaches. The objective of the first use-case scenario is to assist replace a conventional data network while the objective of the second is to create a communication network where no conventional network exists. 

\subsection{Use-case 1: Assisting conventional data networks}


In this first use-case scenario, the ad hoc communication channel is created as a result of the movements of entities equipped with storage capabilities, as depicted in Figure~\ref{fig:dual-channel-assist}. 

This use-case scenario motivates the need to overcome various limitations regarding the coverage and the capacity of conventional data networks. The communication coverage is limited by the high cost of extending or upgrading existing communication infrastructures. This creates a \textit{digital divide} mainly found in areas with challenged environments. Typical environments include rural areas or developing countries. In these scenarios, the digital divide is addressed with public computer kiosks where data is brought by existing mobile entities such as buses to near cities with Internet access~\cite{seth2006low}.

Scenarios focusing on extending the capacity of conventional data networks refer to the emerging paradigm of \textit{traffic offloading}~\cite{patterson2003conversation}. A fraction of traffic is transloaded\footnote{Objects are said to be ``transloaded'' if they are transferred between two different modes of transportation~---~from the Internet to moving entities in this case.} from the data network to another mode of transportation typically involving mechanical vehicles such as cars, airplanes, or trains. The mobile entities can be the vehicles or their passengers carrying data storage devices. As so, data is handled in a similar way as intermodal transportation systems. The objective is to exploit the bandwidth resulting from the combined storage of the mobile entities to help legacy networks such as the Internet cope with the exponential growth of data~\cite{hecht2016bandwidth,index2016forecast}.

Note that the actions of loading data on or off the vehicles may be coordinated by a central controller. Given that these actions need to be taken in a timely manner, the controller communicates with the vehicles via a low-latency legacy infrastructure-based network (\eg cellular network or wired network). Such network is used as an out-of-band channel for the only purpose of exchanging the low-overhead control information in a timely manner. This network enables low-latency communications required by the control messages, contrary to the in-band channel that results in the movements of the vehicles, which is used to offload data traffic from the infrastructure-based networks.


\subsection{Use-case 2: Creating alternative communication networks} 

The second use-case scenario makes no assumptions regarding any existing communication network, as depicted in Figure~\ref{fig:dual-channel-create}. These scenarios seek to create connectivity in challenged environments, such as battlefields or disaster relief, where communication is necessary to support and coordinate military or rescue missions~\cite{zhao2003message}. The objective is to cover areas where the deployment of a communication infrastructure is prohibitively high-cost or challenging due to the scale or the type of environment~\cite{fall2003delay}. 

A recurrent example of such scenario relates to sensor monitoring and collection of environmental data. Mobile entities such as robots collect data from sensors scattered in a target area to monitor~\cite{partan2007survey}. Sensors can either be deployed at fixed locations or mounted on mobile entities. The data is then carried and delivered to a sink for long-term storage and analysis. In the case of a remote distant processing server, the sink acts as a gateway where the data is stored and forwarded to the server using a conventional data network. In this case, the purpose of such scenario is also to extend the coverage limitation of the conventional data network where processing is located.

\subsection{Technical and scientific challenges for data delivery using mobile entities}

The use of mobile entities to deliver data is subject to several challenges and trade-offs. In this section, we highlight the main common challenges shared by the different data delivery approaches. 

The first common challenge is to select the entity best suited for the data delivery, according to the requirements and the environment of the transfers. The selection can be achieved through a characterization of the behavior of the entities. Once an entity is selected, the second challenge is to balance the trade-off between the cost of deploying a specific delivery method and the performance gain brought by the method. For instance, deploying a supporting infrastructure with static intermediate nodes can increase the delivery performance with a higher incurred cost. The third challenge is to quantify the system capacity and its temporal variations, as the movements of the entities are subject to diurnal, weekly, and seasonal variations. The characterization of the system capacity allows  for determining the data transportation resources provisioned by the mobile entities.  The fourth challenge is in allocating these resources in order to maximize their use. For instance, using replication increases the transfer reliability at the expense of capacity. Composing several flows of mobile entities traveling in different directions largely increases the capacity of the data delivery. The challenge is then to calculate the vector route followed by the data as the result of the composition of entity trajectories. This calculation happens on-the-fly as the entity movements are generally not known in advance. Since the selection happens each time two entities encounter each other, choosing an entity that will bring the data away from its destination can decrease the performance of the transfer.

Table~\ref{tab:classification-challenges} summarizes all the data delivery approaches along with their recurring challenges and trade-offs. 


%% file: 1c-mobile-entities.tex
\section{Mobile entities}
\label{sec:entities}

A number of mobile entities have been considered in the literature to carry out the role of data carriers. In this section, we first give a generic model for the alternative communication channel that results from the mobile entity movements. Second, we examine different data transfer scenarios involving entities.

\input{classification-tree.tex}

\input{table_classification_challenges_r2.tex}

\subsection{Mobile entity model}

\begin{figure}[ht]
    \centering
    \includegraphics[width=0.9\columnwidth]{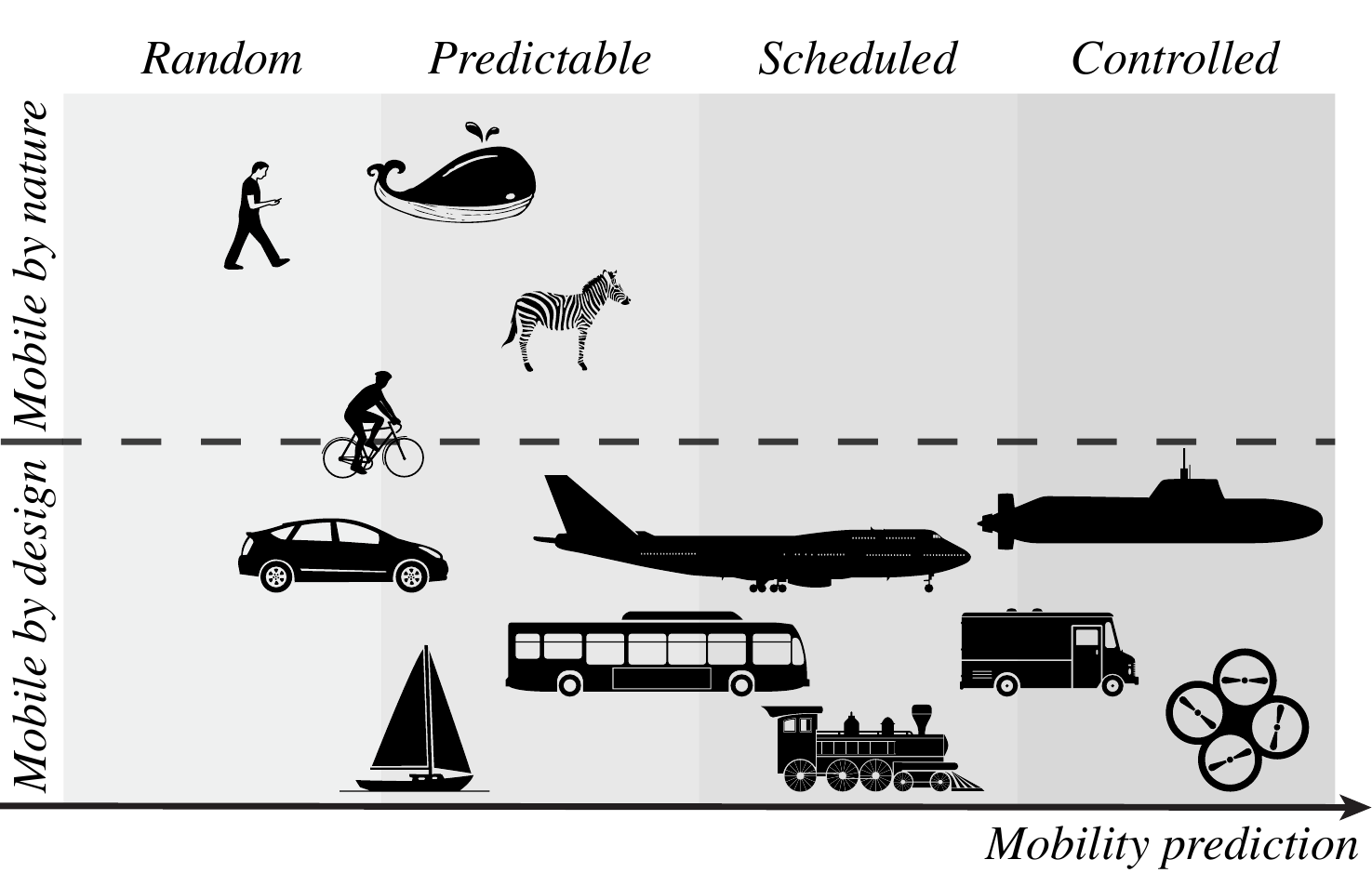}
    \caption{Classification of entities mobile by nature or by design as a function of the degree of predictability of their mobility.}
    \label{fig:mobile-entities}
\end{figure}

A mobile entity can refer to humans, animals, or vehicles equipped with storage devices, such as magnetic disks or other non-volatile solid-state storage devices. As depicted in Figure~\ref{fig:mobile-entities}, the entities are capable of moving either by design (\eg mechanical vehicles powered by combustion or electric engines) or by nature (\eg living beings such as animals or humans carrying devices). They are further equipped with communication capabilities such as wireless interfaces to support data exchange with other entities. The entities may also host the application responsible for producing or consuming data along their trajectories. 

While moving, the entities create a virtual link along which the data stored on the entities' storage devices is transferred. In the case of the direct delivery method, the virtual link connects the source to the destination of the data. In the case of the indirect delivery method, a sequence of mobile entities take turn to deliver the data. As such, the movements of each entity create a virtual link connected to the subsequent one by a virtual router which refers to the data passing location.

\smallskip\noindent\textbf{Virtual links.} The virtual links resulting from the entity movements are analogous to physical links between routers. Unlike physical links where bits are propagated on the wire, the virtual links result from the movements of the entities carrying data. Common network link metrics apply to the virtual links, which can be characterized in terms of bit error rate, throughput, and propagation delay among others. In the case of the virtual links, these metrics depend on the predictability of the entity movements, as well as the number and speed of the entities. The entity movements are generally captured by mobility models we review in Section~\ref{sec:mobility-models}.

\smallskip\noindent\textbf{Virtual routers.} In conventional networks, the purpose of physical links is to connect routers and hosts together to enable end-to-end data transfers. A virtual router refers to the location where the data can be passed either in a synchronous manner when two entities meet together or asynchronously by buffering the data at intermediate stationary nodes called \textit{throwboxes}~\cite{zhao2006capacity}. The output interfaces of a virtual router consist of the multiple entities available for passing the data, and the forwarding decision amounts to the selection of the subsequent entity in charge of carrying the data. The data may also be replicated by passing it to many entities. Two virtual links may be connected together by a virtual router although the trajectories of the underlying entities do not intersect. In this case, the virtual router is the result of the movements of intermediate relay nodes called \textit{ferries}~\cite{zhao2004message} in charge of bridging distant entities.

\subsection{Entity-enabled data transfer scenarios}

Selecting the entities is crucial for the performance of the resulting data transfers. Three types of mobile entities have been considered in the literature we survey: The instrumentalized entities whose movements are opportunistically used to carry data, the controlled entities following precalculated routes, and the contracted entities used by delivery services.  

\smallskip\noindent\textbf{Instrumentalized entities.} This first type of entities refer to those already living (such as humans or animals) or operating (such as public buses) in the target environment. Examples of such entities include whales equipped with sensors that contribute to the collection and the transmission of sensed data~\cite{small2003shared}. In a similar way, zebras have been equipped with collars that log their location history. In this case, zebras are carriers but also the data producers for being the tracked species. Humans have also been considered, such as grape pickers used to collect sensor data in vineyards~\cite{burrell2004vineyard}. Examples of mechanical vehicles include public buses used to bring Internet connectivity to remote rural areas by transporting data as part of their regular service~\cite{pentland2004daknet}. The instrumentalized entities have been referred to as \textit{data mules} and have more or less predictable movements, depending on the entity.   

\smallskip\noindent\textbf{Controlled entities.} The second type of entities refer to trained animals or robots whose trajectories is pre-calculated for the purpose of the data delivery. Referred to as \textit{data ferries}, their use brings the benefit of higher delivery guarantees in terms of delivery delay and reliability. The calculation of the routes followed by the data ferries is challenging. A data ferry may need to visit several nodes where data can be generated at different rates, and can change dynamically, which requires on-the-fly route re-calculations.

\smallskip\noindent\textbf{Contracted entities.} The last type of entities refer to the trucks and airplanes operated by a parcel delivery company contracted to transport the data. If a pickup and delivery can be arranged, the calculation of the routes followed by the parcel are determined by the delivery services and cannot be controlled. The trajectory of the trucks can still be tracked using the online services provided by the delivery company. Given the bandwidth resulting from the number of devices that can be shipped and the geographical coverage, paid parcel delivery service has been considered for assisting conventional data networks such as the Internet or even dedicated lines by offloading large amounts of data.

Figure~\ref{fig:classification-survey} summarizes the classification according to the data delivery approaches we present in the survey. To complement this figure, Table~\ref{tab:classification-challenges} summarizes all the data delivery approaches along with their recurring challenges and trade-offs. Finally, Table~\ref{tab:classification} summarizes the solutions and variations of the different data delivery approaches. Both tables are divided into the direct and indirect delivery approaches.

%% file: classification-tree.tex
\begin{figure*}[t]
\centering
\resizebox{\textwidth}{!}{%
\begin{tikzpicture}[every tree node/.style={align=center,anchor=north},level 1/.style={level distance=1.5cm},level 2/.style={level distance=2cm},level 3/.style={level distance=2cm}]
\Tree [.\textbf{\large Data delivery} 
        [.{\textbf{Direct} data delivery\\(single or multiple entities)\\Section~\ref{sec:direct-delivery}}
            [.\node[text width=3cm]{\textbf{Passive} method\\(existing mobility)\\Section~\ref{sec:direct-non-control}}; 
                [.\node[text width=3.5cm]{\textbf{Random} mobility\\Section~\ref{sec:direct-passive-random}}; ]
                [.\node[text width=3.5cm]{\textbf{Predictable} mobility\\Section~\ref{sec:direct-passive-predictable}}; ] 
                [.\node[text width=3.5cm]{\textbf{Scheduled} mobility\\Section~\ref{sec:direct-passive-scheduled}}; ] ] 
            [.\node[text width=3cm]{\textbf{Active} method\\(controlled mobility)\\Section~\ref{sec:direct-control}}; ] 
            [.\node[text width=3cm]{\textbf{Paying} method\\(postal or courier)\\Section~\ref{sec:contracted-mobility}}; ] ] 
        [.{\textbf{Indirect} data delivery\\(multiple entities)\\Section~\ref{sec:indirect-delivery}} 
            [.\node[text width=4cm]{\textbf{Asynchronous} data passing\\Section~\ref{sec:indirect-async}}; 
                [.\node[text width=3cm]{\textbf{Stationary}\\intermediate nodes\\Section~\ref{sec:indirect-async-stationary}}; ] 
                [.\node[text width=3cm]{\textbf{Mobile}\\intermediate nodes\\Section~\ref{sec:indirect-async-mobile}}; ] ]
            [.\node[text width=4cm]{\textbf{Synchronous}\\data passing\\Section~\ref{sec:indirect-sync}}; 
                [.\node[text width=3cm]{\textbf{Floating}\\composition\\Section~\ref{sec:indirect-sync-non-anchored}}; ]
                [.\node[text width=3cm]{\textbf{Pre-positioned}\\composition\\Section~\ref{sec:indirect-sync-anchored}}; ] ] ] ]

\end{tikzpicture}}
\caption{Classification of the methods proposed for transferring data using one or many mobile entities according to the data delivery approaches (direct or indirect). Each delivery approach is further classified depending on the type of mobility (random, predictable, or scheduled) when a single mobile entities are involved and on the methods used to pass the data (asynchronous or synchronous) if many relay mobile entities are used. We also give the sections surveying the corresponding approaches.}
\label{fig:classification-survey}
\end{figure*}

%% file: table_classification_challenges_r2.tex
\begin{table*}[t]
  \small
  \centering
    	\caption{Recurrent challenges and tradeoffs for the different data delivery approaches in the classification.}
  \resizebox{1.8\columnwidth}{!}{%
  \begin{tabular}{p{0.5cm}L{4cm}L{6cm}L{5cm}} 
  \toprule  
  \multicolumn{2}{l}{\textbf{Data delivery approach}} & \textbf{Challenges} & \textbf{Trade-offs}\\
  \midrule  
  
  \multirow{3}{*}{\rotatebox[origin=c]{90}{\parbox[c]{1.5cm}{\centering Direct data delivery}}} & Passive & Logistics at endpoints and which entity to use & Data delivery ratio and delay vs. cost\\[3pt]

 & Active & Static route and schedule planning for the controlled entities & Delivery delay and ratio vs. cost and energy efficiency \\[3pt]
 
 & Paying & Cost planning to contract the service & Delivery delay and throughput vs. cost \\
 
 \cmidrule(l){1-4} 
 
 \multirow{4}{*}{\rotatebox[origin=c]{90}{\parbox[c]{2.5cm}{\centering Indirect data\\delivery}}} & Asynchronous data passing (stationary relay nodes) & Optimal relay node placement (load sharing, buffer management) & Number of nodes deployed and capacity vs. cost \\[3pt]
 
 & Asynchronous data passing (mobile relay nodes) & Dynamic route and schedule planning of multiple entities & Number of mobile nodes vs. delivery delay and cost \\[3pt]
 
 & Synchronous data passing (floating locations) & When to pass or replicate data from one node to another & Protocol overhead vs. data delivery ratio and delay \\[3pt]
 
 & Synchronous data passing (pre-positioned locations) & Determination of the pre-positioned passing locations & Protocol overhead vs. data delivery ratio and delay\\
	 
 \bottomrule

  \end{tabular}
  }
  	\label{tab:classification-challenges}
\end{table*}

%% file: 3a-direct-delivery.tex
\input{table_classification.tex}

\section{Direct data delivery overview}
\label{sec:direct-delivery}

In this section, we present a classification of the approaches that use a direct delivery approach: a single entity or a group of independent entities are in charge of delivering data between a source and the final destination. The approaches we review in this section are listed in the left column of Table~\ref{tab:classification}. The benefits of the direct delivery approach increase with the predictability and the resulting knowledge regarding the entity mobility. This has motivated the choice of specific kinds of entities and the careful deployment of the data sources and destinations, if stationary.


\begin{itemize}
 
 \item \textbf{The passive method} (Section~\ref{sec:direct-non-control}) opportunistically exploits the existing mobility of a single entity referred to as a data mule. The performance of the passive method depends on the degree of knowledge of the mobility of the entity which may result from the predictability of their movements. 
 
 \item \textbf{The active method} (Section~\ref{sec:direct-control}) consists of controlling the mobility of entities commonly referred to as data ferries . These entities follow precalculated routes with the purpose of improving the data delivery rate and delivery latency.

 \item \textbf{The paying method} (Section~\ref{sec:contracted-mobility}) uses the paid services of a postal or delivery company. This method is a combination of the two previous methods: the delivery trucks may be considered as controllable entities, nevertheless the calculation of their routes is determined by the delivery services. Furthermore, the use of this mobility results in an active method since delivery services are purchased with the obvious purpose of transporting data.
 
\end{itemize}

We first review the passive method which can be leveraged to create an ad hoc communication channel if entities are already moving in the target area. Second, we review the active method that can be used if no entities are available in the area or to ensure reliable data transfers. Finally, we review the paying method useful to cover long distances.

\section{Passive method for direct data delivery}
\label{sec:direct-non-control}

\begin{figure}[t]
    \centering
    \includegraphics[width=0.85\columnwidth]{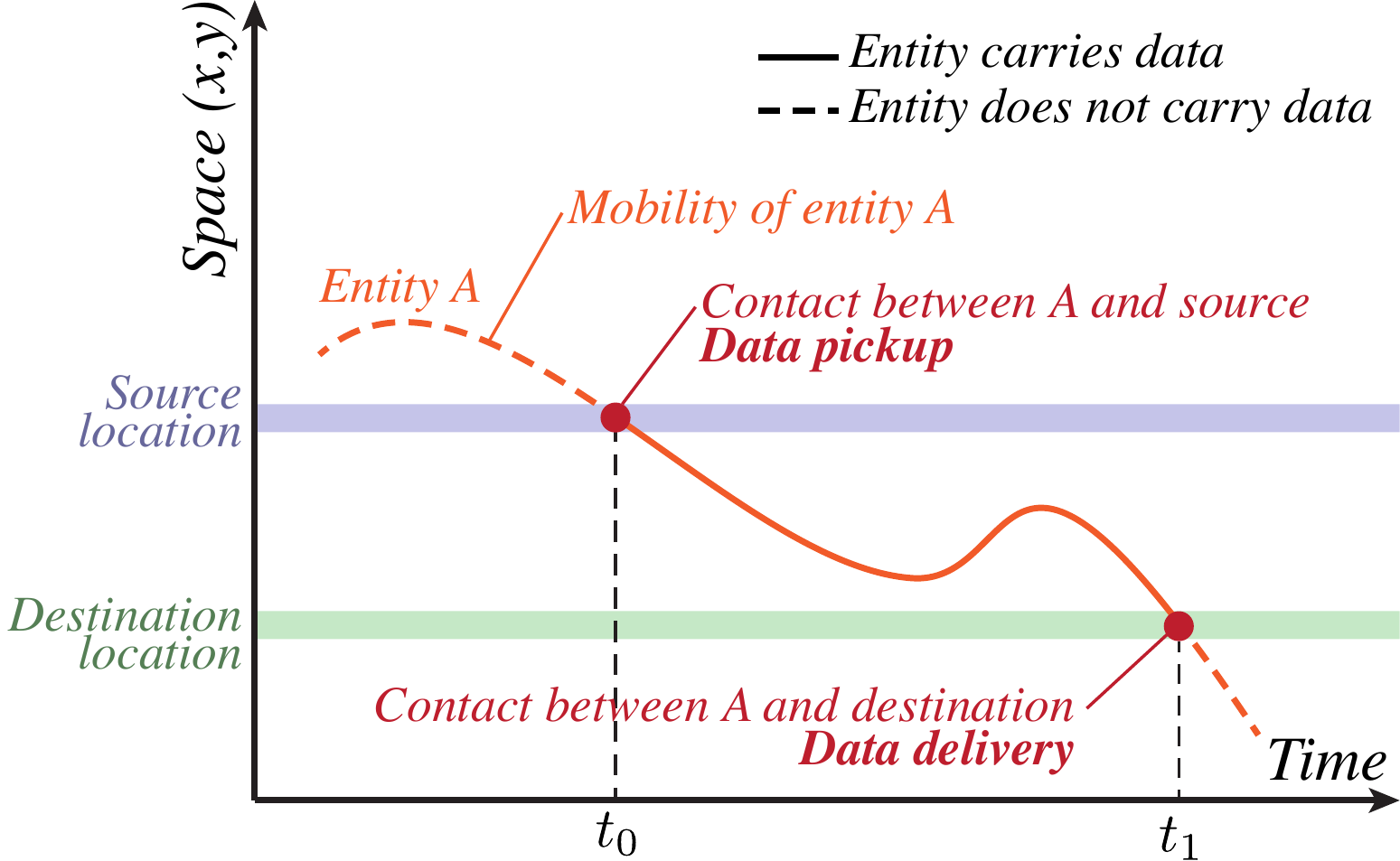}
    \caption{Passive method for direct data delivery using a single entity. The entity $A$ picks up the data from the source when in the source's transmission range at time $t_0$ and carries it to the destination, where $A$ delivers the data at time $t_1$.}
    \label{fig:dtn-direct-forwarding}
\end{figure}

Relevant applications using the passive method include sensing platforms where entities such as boats or buses are used to create connectivity between sensors deployed in areas not covered by conventional data networks. Referred to as data mules in~\cite{pentland2004daknet}, these entities can help bridge connectivity gaps between sources and destinations (\eg sensor nodes and sinks for sensing platforms). The passive method for direct data delivery is depicted in Figure~\ref{fig:dtn-direct-forwarding}. In this figure, a single entity $A$ carries data it picked up when in contact with the data source at time $t_0$ (\eg when in the transmission range of a sensor node) to the data destination at time $t_1$ (\eg when in the transmission range of a sink). The existing mobility of entity $A$ enables the data transfer.

\subsection{Random mobility}
\label{sec:direct-passive-random}

The premises of the passive method were first introduced by Grossglauer and Tse~\cite{grossglauser2001mobility}. The objective was to evaluate the performance of this method in comparison to a MANET network. Their results show that under high data loads, the random mobility of an entity carrying data while moving helps achieve better throughput. They also confirmed that the use of a mobile entity consumes less energy and bandwidth by removing the need of relaying data from one node to another. 

The passive method can be seen as a degenerate case of the First Contact strategy~\cite{jain2004routing}. In this strategy, a predefined number of independent entities receive a copy of the data when in contact with the source. The data is then eventually delivered if at least one of these entities acting as a relay node comes in contact with the destination. The First Contact strategy is similar to the passive method if used with a single intermediate entity.

The Data MULE (Mobile Ubiquitous LAN Extension) architecture was proposed in the context of sensing platforms~\cite{shah2003data}. Data MULEs refer to mobile entities such as people, animals, or vehicles whose movements are instrumentalized to carry data collected from sensors scattered in a sparse area and delivered to access points acting as gateways. 
 
Despite its simplicity, the main drawback of the passive method used with a single entity moving at random is the high delivery latency which results from the time it may take for the entity to come in contact with the destination, if ever.

\subsection{Predictable mobility}
\label{sec:direct-passive-predictable}

Instead of relying on entities moving at random, the following works make use of various entities including humans, boats, or buses whose mobility can be predicted to some extent. These entities move along routes whose endpoints are known in advance. In the context of a sensing platform, knowing the route endpoints allows the informed placement of the data sinks.       

Burrell~\etal propose to use the mobility of grape pickers to collect environmental data from a collection of sensors deployed in a vineyard~\cite{burrell2004vineyard}. The sensor data is then gathered at a sink located in the farm where workers make compulsory stops as part of their work routine. The centralization of the sensor data helps take effective decision in face of low temperatures, rain, or frost. 

SeNDT (Sensor Networking with Delay Tolerance) proposes the design of a sensor platform for challenging environments~\cite{mcdonald2007sensor}. This platform was deployed in the context of two applications including lake water quality monitoring. The sensing platform relies on the routes taken by angler boats while fishing to collect the data from anchored sensors monitoring the water quality of lakes in Ireland. The angler boats communicate with the sensors via standard WiFi communications and deliver the collected data once they return to the boathouse. We present the second application in the following section.

In addition to sensor data collection, several works have also proposed to leverage the existing flows of mobile entities (\eg people with smartphones or cars) for data transfers between two endpoints. This is the case of Mansy~\etal that exploits the flows of cars on a highway to deliver data between distant roadside units~\cite{mansy2007reliable}. The authors focused on the reliable data delivery by adapting well-known retransmission algorithms (\eg ARQ) to the architecture. In another application context, Hoseini~\etal leverage the predictable movements of mobile devices to provide a low-cost data transportation option for cellular backhaul~\cite{hoseini2014message}. A continuous flow of mobile devices pick up data at small cells (deployed at traffic hot spots) and deliver it to switching centers. The dual flow of mobile devices also exists in the opposite direction. The authors mainly focused on the memory management issues related to the limited storage availability in smartphones. Their solution is an algorithm that predicts the available storage of smartphones over a 7.5 minute period with 94\% accuracy.

\subsection{Scheduled mobility}
\label{sec:direct-passive-scheduled}

Scheduled mobility refers to the regular routes taken by vehicles such as the buses operated at specific hours by the public transit agencies. This is a special case of predictable mobility which provides the passive method with a higher level of knowledge of the instrumented vehicles' mobility.

In DakNet, buses serving remote rural areas of India and Cambodia are used as a cost-effective replacement for expensive dialup landlines or long-range radios~\cite{pentland2004daknet}. The regular schedule of the buses helps extend the Internet connectivity of larger cities to those areas. The buses are equipped with wireless communication devices and collect the delay-tolerant data such as emails or land records from public computer kiosks. The data are then brought to the city and offloaded to wireless access points acting as Internet gateways. 

A similar initiative was proposed by the Wizzy Digital Courier service to provide Internet connectivity to schools located in remote villages of South Africa~\cite{rabagliati2004wizzy}. As with Daknet, the regular schedule of buses tends to improve the performance of the passive method.  

In the EMMA project, environmental sensors are mounted directly on the public buses or trams for monitoring the air pollution in cities~\cite{lahde2007practical}. The high mobility and density of public transportation networks are exploited so to provide accurate and up-to-date measurements. The mobile sensors offload their data at gateways strategically deployed where vehicles pass by frequently, such as intersections. The sensor data are then forwarded to a sink (e.g. the evaluation server) via a conventional infrastructure-based data network.

The second application of the SeNDT platform already introduced in previous section, targets noise level monitoring in urban areas and on motorways~\cite{mcdonald2007sensor}. The noise sensors are mounted on traffic light poles and the scheduled routes of garbage or delivery trucks are instrumented to collect and deliver the sensed data once at their depot.

\section{Active method for direct data delivery}
\label{sec:direct-control}

\begin{figure}[t]
    \centering
    \includegraphics[width=0.85\columnwidth]{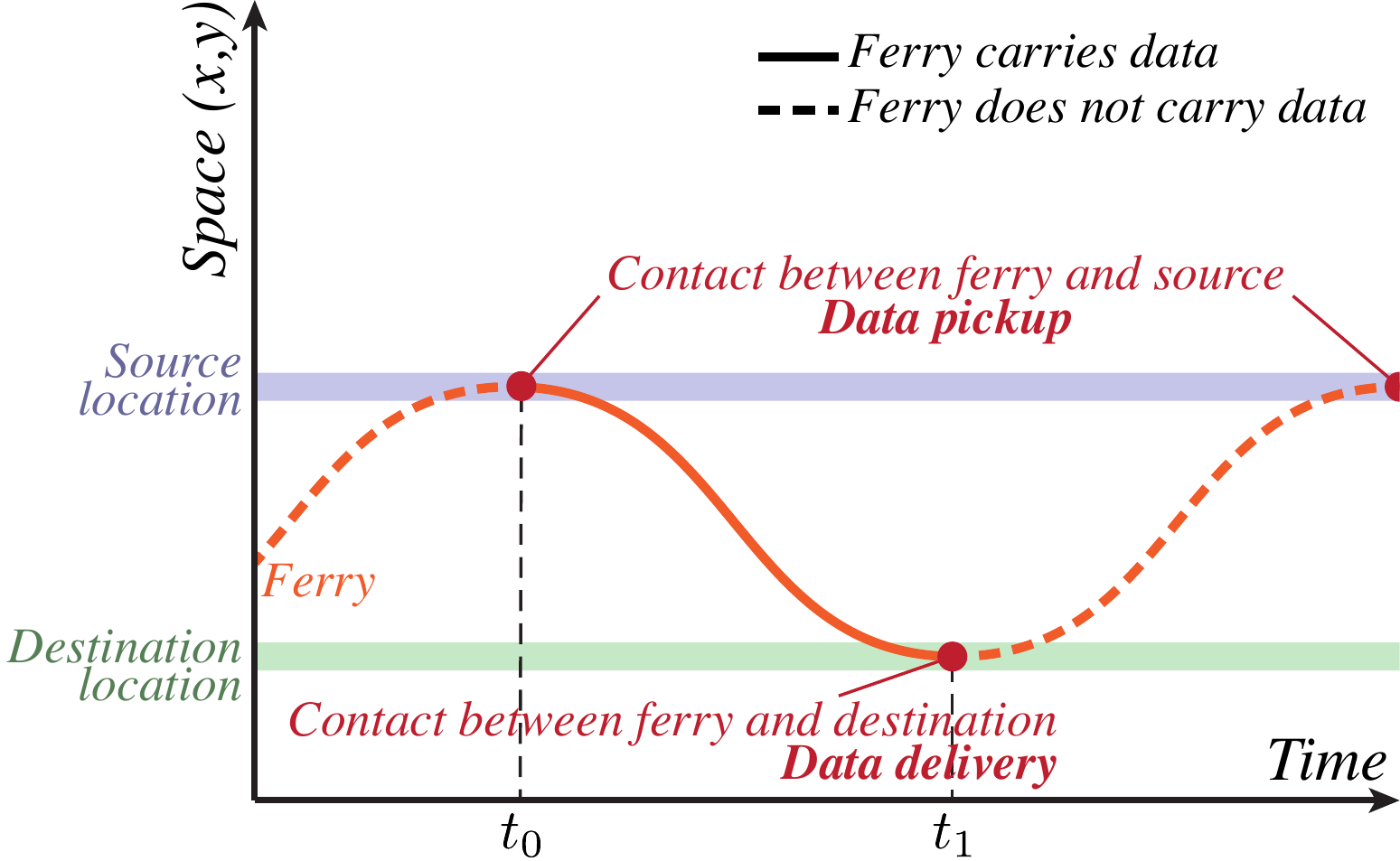}
    \caption{Active method for direct data delivery with a single controlled entity. The ferry picks up the data at the source at time $t_0$, carries it, and delivers it to the destination at time $t_1$.}
    \label{fig:dtn-direct-controlled-forwarding}
\end{figure}

Instead of using the existing mobility of instrumented entities, the active method consists in controlling the mobility of entities such as trained animals or robots. With this method, the mobile entities actively modify their trajectories to transmit and deliver the data they carry~\cite{li2000sending}. Commonly referred to data ferries, these entities follow precalculated routes with the purpose of improving the data delivery rate and delivery latency. Figure~\ref{fig:dtn-direct-controlled-forwarding} depicts a use-case scenario where the route followed by a ferry was calculated so as to periodically visit the source and destination locations represented by the blue (above) and green (below) bands, respectively. The ferry's precalculated movements enable the data transfer. 

A first example of the active method is provided in RFC 1149~\cite{rfc1149} (and its later improvements). Originally proposed to illustrate (in a humorously way) the mantra:  ``IP over everything'', this RFC presents an experimental method for carrying IP packets on avian carriers (IPoAC) such as homing pigeons. The latter have been long bred for their ability to find their way back home over extremely long distances. Training homing pigeons involves mentally marking specific points as their home and has enabled their use as message carriers in various situations such as wars or more regular services including pigeon post or for emergency communication following natural disasters.

The main challenge addressed by the approaches using the active method lies in the calculation of the routes followed by the ferries. This calculation is driven by the requirements regarding the locations to be visited besides the final destination. In the work we review, the calculated routes are also characterized by the bandwidth requirement given the rate (uniform or not) at which the data is generated at those locations. If the bandwidth resulting from the entities having taken those routes is insufficient, the data accumulates until dropped in case of locations with limited buffer capacity. Common performance metrics for the calculated routes include the data delivery delay and delivery rate. 

\subsection{Single controlled entity}
\label{sec:single-entity}

Zhao \etal introduced the message ferrying scheme where message ferries are in charge of carrying messages between a set of disconnected nodes, some acting as senders while others as destinations~\cite{zhao2003message}. They propose a ferry route calculation algorithm which objective is to minimize the delivery delay of the collected data. This algorithm assumes that data is generated by all sending nodes at the same rate. The algorithm first creates a route using delay-based local optimization techniques to adapt the existing Travel Salesman Problem (TSP) algorithms so as to minimize the average delivery delay of the data collected from all sending nodes. The algorithm then extends the original route to meet the bandwidth requirements resulting of the nodes' data generation rate. The route extension consists in calculating a detour that increases the time ferries spent in the vicinity of a node. 

Mansy \etal consider nodes with different data generation rates which results in varying bandwidth requirements~\cite{mansy2011deficit}. They use the deficit round-robin technique to determine in which order to visit the nodes depending on the rate at which they generate data. With this approach, the likelihood of visiting a node increases with the node data rate. 

Tirta \etal propose to decrease the delay of the ferry route by grouping the nodes in clusters~\cite{tirta2004efficient}. Data is first locally gathered at specific nodes called cluster heads before being collected and distributed to all clusters by visiting ferries referred to as data collectors. The route of the collectors across the cluster heads and the frequencies at which a cluster head is visited, are calculated according to various visit schedule policies. 
The authors show that a schedule policy based on both the cluster heads' data generation rates and the distance traveled by the collector helps reduce the energy consumption of the data collector, whereas a schedule policy based only on the aggregated data generation rate leads to better average data collection latency.

\subsection{Multiple controlled entities}

\begin{figure}[t]
    \centering
    \includegraphics[width=0.85\columnwidth]{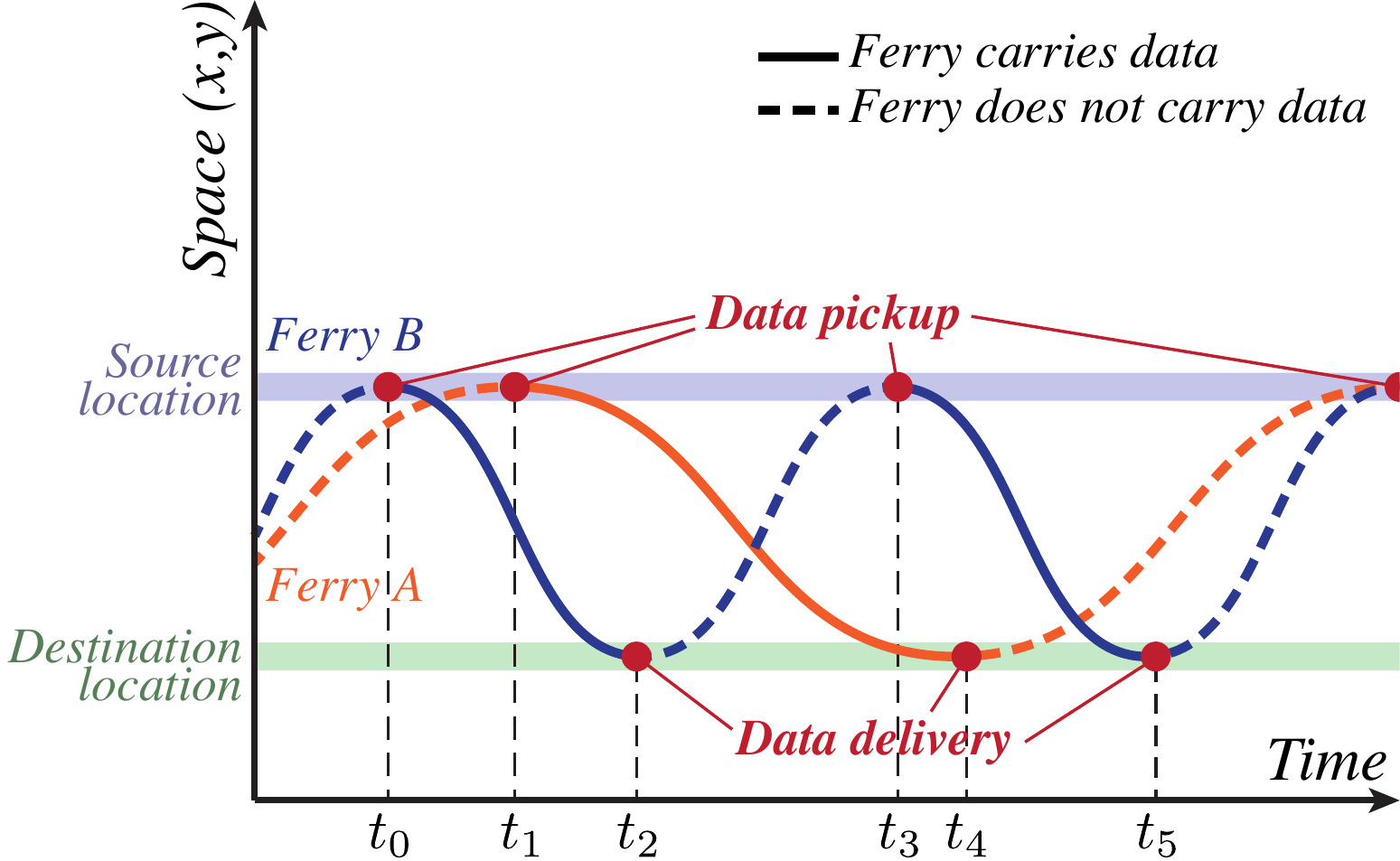}
    \caption{Active method for direct data delivery with multiple controlled entities.}
    \label{fig:dtn-direct-controlled-forwarding-multiple}
\end{figure}

Instead of relying on the route calculated for a single entity, the combined actions of a fleet of independent controllable entities is expected to bring performance improvements in terms of delivery delay and ratio resulting from the cumulative distance covered or data carried by all entities. Figure~\ref{fig:dtn-direct-controlled-forwarding-multiple} depicts a fleet of two ferries $A$ and $B$ that periodically transfer data from the source to the destination at different rates: Ferry $B$ delivers data twice as fast as Ferry $A$. Relying on multiple controlled entities provides robustness in face of entity failure or malicious attacks.

\subsubsection{\textbf{Data transport ferries}} Zhao \etal propose two algorithms for calculating the routes of ferries operating together to carry data among a collection of nodes, some acting as the sources, others as the destinations of the data~\cite{zhao2005controlling}. The first algorithm is similar to the one they proposed for the case of a single controlled ferry in~\cite{zhao2003message} (see Section~\ref{sec:single-entity}). This algorithm calculates a single route followed by all the ferries moving at same speed but with different timings and directions. The route is calculated so each ferry visits all nodes, allowing each of them to communicate with all the ferries. In a second algorithm, multiple routes are calculated each followed by different ferries. 

In both algorithms, the route calculation follows the same steps: First, each ferry is assigned to a pair of source-destination nodes and the route connecting this pair is calculated by minimizing its length expressed as the number of ferry hops. This metric captures the waiting time before the source receives the visit of the ferry and the delivery delay needed by the ferry to reach the destination. In a second step, the route is refined so as to match the bandwidth requirements defined by the nodes' data generation rates given the capacity of each route resulting of the number, speed, and cargo size of the ferries. 

\subsubsection{\textbf{ Pigeon networks}} More recently, a special case of ferries referred to as pigeons was studied by Zhou~\etal~\cite{zhou2013minimizing}. A pigeon corresponds to a special-purpose vehicle and has the particularity of being dedicated to a specific node called the home host. A pigeon carries and delivers messages from its home host to the rest of the nodes, called the foreign hosts, before returning to its home host. The authors propose various optimization methods for the calculation of a single pigeon route that minimizes the average delivery delay of a message. Each method is designed for specific scenarios depending on the number of foreign nodes. 

For low foreign node densities, the calculated route is optimal whereas its calculation becomes intractable for higher densities. They address high densities by using a geographical partitioning-based heuristic which relies on a divide-and-conquer approach to design the pigeon route. The authors show that the resulting routes allow lower delivery delays compared to a non-partitioning approach similar to the route calculation introduced in the message ferry work of Zhao~\etal~\cite{zhao2003message}.

\subsection{Real-world deployments and experiments}

Real-world experiments have been conducted in the context of sensing platforms with the purpose of studying the benefits of deploying controllable entities. 

\subsubsection{\textbf{Terrestrial sensor networks}} Tekdas~\etal deployed a small-scale wireless sensor network to measure the energy savings brought by the addition of a fleet of robots~\cite{tekdas2009using}. Each robot is assigned to a subset of sensors they visit by taking a TSP (Travel salesman problem) tour. Once their tour is completed, the robots return to the gateway where they offload the data they have collected. The calculation of the tours for $k$ robots consists first in finding the complete optimal TSP tour covering all sensors. The k-SPLITOUR algorithm is then used to split the TSP tour in $k$ smaller tours~\cite{frederickson1976approximation}. A small-scale network consisting of twelve sensors and one gateway was deployed on a basketball field. The results of three experiments with a number of robots varying from 0 to 3 show that the use of robots significantly reduces the sensors' energy consumption. The ability of a robot to move close to the sensors allows them to reduce their transmission power level while improving the quality of the wireless link. By removing the need of ad hoc hop-by-hop transmissions between neighboring sensors, the robots contribute to prolong the sensor network lifetime.

\subsubsection{\textbf{Underwater sensor networks}} Underwater wireless sensor networks provide another example of a real-world deployment of robots referred to as Autonomous Underwater Vehicles (AUVs)~\cite{partan2007survey,dunbabin2006data}. AUVs can serve a wide range of tasks including collecting data from underwater sensors deployed in the oceans for long-term environmental monitoring or surveillance. Because of the cost of the technology embedded on such sensors and the extent of the areas targeted by the monitoring, underwater wireless sensor networks are usually sparser than terrestrial sensor networks. Another difference results from the challenges of underwater communications which motivate the use of acoustic channels. Such a channel poses severe issues regarding the navigation system of the AUVs. Navigation and communication signals usually share the same frequency bands. The resulting contention leads to navigation errors. In~\cite{dunbabin2006data}, the AUVs have a map of the sensor locations and navigate to the next closest sensor using visual odometry for data collection. The AUVs use an optical high-bitrate channel for data transfer and short communication with the sensors. A set of 46 experiments were conducted with three sensors deployed in a pool to assess the AUV visual-based navigation system and demonstrate the ability of AUVs to visually locate and dock with up to 8 sensors.

\section{Paying method for direct data delivery}
\label{sec:contracted-mobility}

\begin{figure}[t]
    \centering
    \includegraphics[width=0.85\columnwidth]{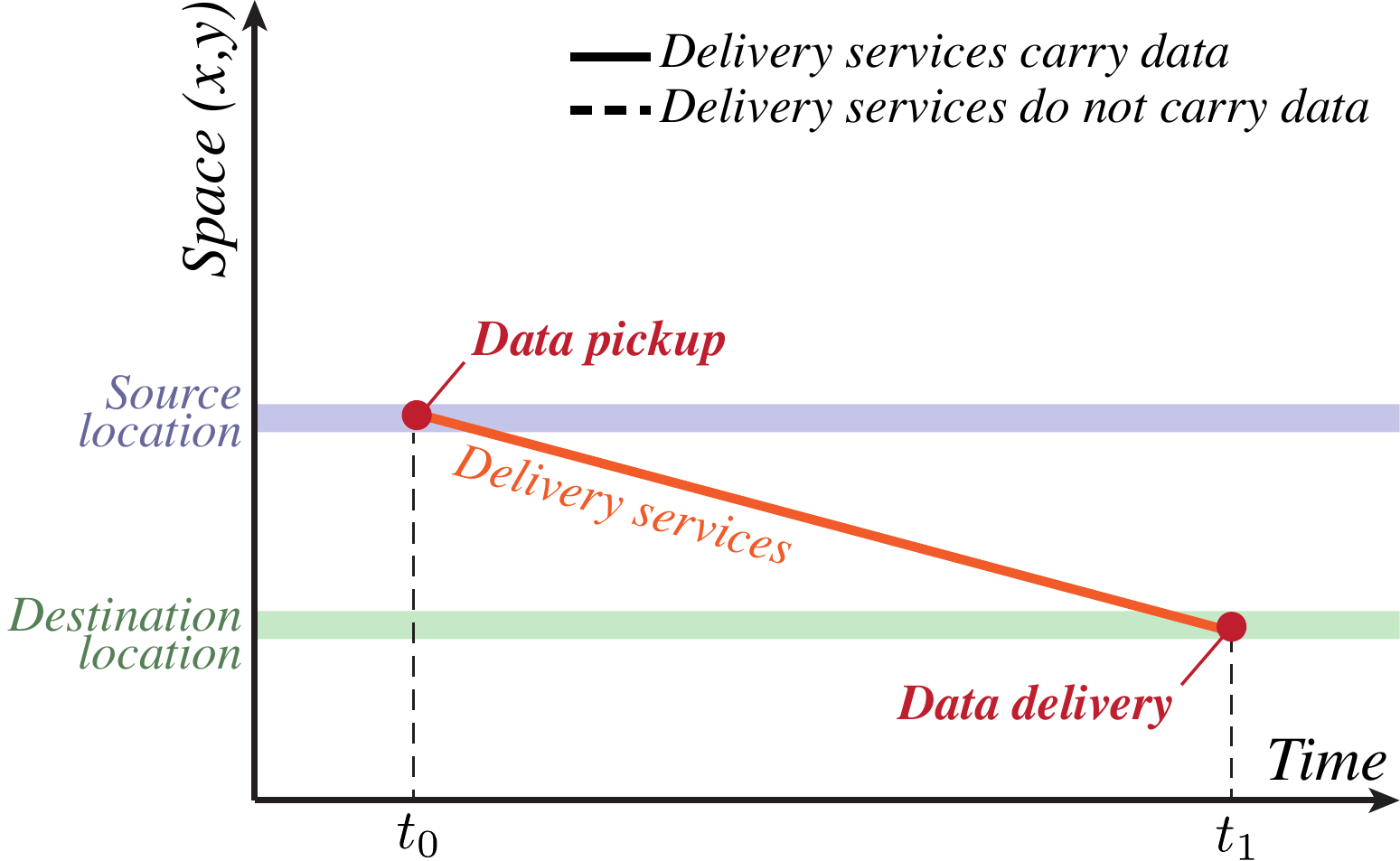}
    \caption{Paying method for direct data delivery using delivery services.}
    \label{fig:dtn-direct-delivery}
\end{figure}

In this section, we review the approaches that rely on the paid services of a postal or delivery company. The trucks operated by such companies are special cases of data ferries since delivering is their primary use. Data is stored on memory storage media such as hard drives or disks seemingly packed into shipping boxes. The delivery vehicles may be considered as controllable entities, nevertheless the calculation of their routes is integrated to the service provided by the delivery company. Furthermore, this mobility results of an active method as delivery services are purchased with the stated purpose of transporting data. We represent their use in Figure~\ref{fig:dtn-direct-delivery} where postal or courier services directly transport the data from its source location (shown in blue above) to its destination location (shown in green below).

\subsection{Internet improvements}
\label{sec:internet-improv}

The postal or courier services have been considered with the purpose of bridging the connectivity gap of rural areas or offloading the Internet from large amounts of traffic.

\subsubsection{\textbf{Extending Internet connectivity}} Wang \etal propose Postmanet, a system designed to extend the Internet connectivity to rural areas or in developing countries by turning the postal services into a generic transmission medium~\cite{wang2004turning}. Postmanet relies on public kiosks acting as receptacles for outgoing or incoming mobile storage media such as DVDs. On the user side, the kiosks are provided with slots where users can insert or retrieve DVDs as if using a standard computer. On the postal service side, the kiosks act as a mailbox where DVDs are collected or delivered as standard mail is. The design of the Postmanet system is concerned with transport-level issues such as DVD damages or losses, delayed or out of order deliveries. To address these issues, the Postmanet exploits the simultaneous use of the Internet or the telephone system to send or speak out loud ``out of band'' control messages such as acknowledgements. 

The authors introduce multiple routing strategies including the direct delivery between the Postman kiosks. The postal service is paid for the end-to-end delivery between pairs of kiosks. They also consider various indirect strategies including a peer-to-peer routing approach where a package is sent through a sequence of kiosks before reaching the final destination. Each visited kiosk adds their memory device to the parcel before passing it. We will introduce the other routing strategies in Section~\ref{sec:indirect-async-stationary} that is dedicated to the indirect delivery approach.

\subsubsection{\textbf{Enhancing Internet capacity}} 
Laoutaris~\etal propose a scheduling policy for delaying the transfers of bulk delay-tolerant data~\cite{laoutaris2013delay}. This policy exploits the available off-peak transmission capacity resulting from the diurnal traffic pattern combined to the percentile-based pricing scheme that makes this capacity more affordable. The data is first uploaded to transit storage nodes where they are stored before transferred according to the scheduling policy. To assess the cost benefit of this policy, they compare the price an Internet Service Provider would charge  (using the percentile pricing) to the cost of offloading the same amount of data using the services of a package delivery company (\ie FedEx or UPS). The authors found that using the services of a package delivery company to offload large amounts of traffic off the Internet is less expensive for punctual individual shipments (\eg short-lived transfers). However, in the case of a constant flow of data, they show that the package delivery solution is more expensive than sending it over the Internet. This improvement comes at the cost of further delaying the data transfers using intermediate storage nodes where data is buffered.

A concrete example of traffic offloading is provided by scientific instruments such as CERN's Large Hadron Collider (LHC) or NASA's Hubble Space Telescope (HSN) that generate tens of terabytes or petabytes of data sent to remote collaborators or computational capable centers by parcels~\cite{patterson2003conversation}.

\subsection{Commercial uses}

Many companies rely on postal or courier services to deliver software and digital media to their customers. 

\subsubsection{\textbf{Digital media and software}} An example of such company is Netflix\footnote{Netflix DVD service at \url{http://dvd.netflix.com/}} that started its business by providing a DVD-by-mail rental service before streaming media. While shipping DVDs via mail was the core business model of Netflix before 2011 and the introduction of the streaming service, the service still has 3.5 million customers as of 2017.\footnote{\label{note:netflix-ir}Neflix Investor Relations at \url{http://ir.netflix.com}} Netflix DVD service is complementary to the streaming service, as it allows offloading massive amounts of data and reaching customers without the sufficient capacity to stream videos. This legacy service further provides as alternative way for Netflix to distribute videos when these are not available for streaming (probably due to copyright reasons) by making them available through the DVD service. In 2006, Netflix was distributing more than 1.5~million DVDs a day\footnote{See footnote~\ref{note:netflix-ir}.}, accounting for a total aggregate bandwidth of more than 650~Gbps. Another example of company is AOL who used to send their installation software on floppy disks and later on CD-ROMs via unsolicited mail as part of their direct marketing campaigns.

\subsubsection{\textbf{Cloud computing platforms}} More lately, large cloud computing platforms provide an example of Internet service using the postal or courier services as an offloading channel. Users of those platforms can upload their data by sending memory storage devices such as hard drives via mail or courier. 
Such platforms include AWS Import/Export\footnote{AWS Import/Export service at \url{http://aws.amazon.com/importexport}}, Microsoft Azure Import/Export Service\footnote{Microsoft Azure Import/Export service at \url{http://azure.microsoft.com/en-us/documentation/articles/storage-import-export-service}} and more recently Google Offline Disk Import\footnote{Google Cloud Offline Media Import / Export service at \url{https://cloud.google.com/storage/docs/offline-media-import-export}}). In 2015, Amazon launched the AWS Import/Export Snowball solution which consists of portable ready-to-be-shipped appliances with a storage capacity of 50~TB or petabyte-scale data transport in and out the AWS cloud platform. A similar service is provided by IDrive\footnote{IDrive Express service at \url{https://www.idrive.com/idrive-express}}, a company that provides a backup service to individuals or corporations including the provision of hard drives and the delivery by courier for data uploading or retrieval.

%% file: table_classification.tex
\begin{table*}[t]
\small
  \centering
    	\caption{A classification of data delivery strategies, along with their research directions and surveyed works. }
  \vspace*{-15pt}
  \resizebox{2\columnwidth}{!}{%
  \begin{tabular}{l||lccc} 
  & & & & \\[1.5ex]
  \toprule  
  \textbf{Direct data delivery} &  \multicolumn{4}{c}{\textbf{Indirect data delivery}}\\

  \cmidrule(l){3-5}  
    &  & \textit{Asynchronous delivery} & & \textit{Synchronous delivery}\\  
  
  \midrule
  
   \cellcolor[gray]{0.97} \textit{Passive method} & \parbox[t]{1mm}{\multirow{6}{*}{\rotatebox[origin=c]{90}{\textit{Stationary}}}} & \cellcolor[gray]{0.92} &  \parbox[t]{2mm}{\multirow{6}{*}{\rotatebox[origin=c]{90}{\textit{Floating}}}}  &  \cellcolor[gray]{0.97} \raisebox{-0.7ex} {No control plane} \\
   \cellcolor[gray]{0.97} {Random: \cite{grossglauser2001mobility}, \cite{jain2004routing}} & &  \cellcolor[gray]{0.92} Existing mobility &  & \cellcolor[gray]{0.97} \raisebox{-0.1ex} {
   \cite{small2003shared},
   \cite{jain2004routing},
   \cite{shah2003data},
   \cite{vahdat2000epidemic}, \cite{papdopouli2000seven}, \cite{papadopouli2001design}, \cite{hull2006cartel}, \cite{eisenman2009bikenet}, \cite{goodman1997infostations},  \cite{leontiadis2007geopps}, \cite{soares2014geospray}} \\ 
   \cellcolor[gray]{0.97} {Predictable: \cite{burrell2004vineyard}, \cite{mcdonald2007sensor}, \cite{lahde2007practical}} & & \cellcolor[gray]{0.92} \cite{seth2006low},
   \cite{zhao2006capacity}, \cite{chawathe2006inter}, \cite{ibrahim2009analysis},  \cite{demmer2007dtlsr}, \cite{chen2015dtn}, \cite{zarafshan2010trainnet}, \cite{banerjee2007energy} & & \cellcolor[gray]{0.97} \raisebox{-0.7ex} {Local control plane} \\
  \cellcolor[gray]{0.97} {Scheduled: \cite{pentland2004daknet}, \cite{rabagliati2004wizzy}, \cite{rfc1149}} & & \cellcolor[gray]{0.92} Controlled and paying mobility
  & & \cellcolor[gray]{0.97} {\cite{davis2001wearable},
  \cite{juang2002energy},
  \cite{spyropoulos2005spray}, \cite{spyropoulos2007spray}, \cite{burns2005mv},  \cite{hui2011bubble}} \\ 
  \cellcolor[gray]{0.94} \textit{Active method} & & \cellcolor[gray]{0.92} \raisebox{0ex} {\cite{wang2004turning}, \cite{cho2011budget}, \cite{zhao2005controlling}}  & & \cellcolor[gray]{0.97}  In-band global control plane\\
   \cellcolor[gray]{0.94} {Single: \cite{zhao2003message}, \cite{mansy2011deficit}, \cite{tirta2004efficient}} & & \cellcolor[gray]{0.92}  & & \cellcolor[gray]{0.97} \raisebox{0.9ex} {\cite{lindgren2003probabilistic}, \cite{musolesi2005adaptive}, \cite{costa2008socially}, \cite{balasubramanian2010replication}, \cite{burgess2006maxprop}, \cite{hui2011bubble}} \\
      
      
   \cellcolor[gray]{0.94} \raisebox{-0.7ex} {Multiple: \cite{zhao2005controlling}, \cite{zhou2013minimizing}, \cite{tekdas2009using}} & \parbox[t]{2mm}{\multirow{6}{*}{\rotatebox[origin=c]{90}{\textit{Mobile}}}}  &  \cellcolor[gray]{0.97} \raisebox{-0.7ex} {Centralized global control plane} & \parbox[t]{2mm}{\multirow{6}{*}{\rotatebox[origin=c]{90}{\textit{Pre-defined}}}} & \cellcolor[gray]{0.92} \raisebox{-0.7ex} {In-band global control plane}\\
   \cellcolor[gray]{0.94}  & & \cellcolor[gray]{0.97} \raisebox{-0.3ex} {\cite{zhao2004message}~(NIMF and FIMF),  \cite{bin2006message}} & & \cellcolor[gray]{0.92} \raisebox{-0.3ex} {\cite{lee2008louvre}, \cite{sarafijanovic2006island}, \cite{yuan2009predict}}\\ 
   \cellcolor[gray]{0.92} \raisebox{-0.8ex}{\textit{Paying method}} & & \cellcolor[gray]{0.97} \raisebox{-1.8ex}{Distributed global control plane} & & \cellcolor[gray]{0.92} \raisebox{-1.8ex} {Out-of-band global control plane}\\ 
   \cellcolor[gray]{0.92} \raisebox{0.7ex}{Postal services: \cite{wang2004turning}, \cite{laoutaris2013delay}} & & \cellcolor[gray]{0.97} \raisebox{-0.1ex} {\cite{burns2008mora}, \cite{asadpour2016route}} & & \cellcolor[gray]{0.92}  \raisebox{-0.1ex} {\cite{zhao2005controlling}, \cite{tan2014vehicular}, \cite{keranen2009dtn}}\\ 
   \cellcolor[gray]{0.92} {Commercial: Netflix, AWS} & & \cellcolor[gray]{0.97} & &  \cellcolor[gray]{0.92} \\
	 
  \bottomrule 

  \end{tabular}
  }
  	\label{tab:classification}
\end{table*}

%% file: 3b-indirect-async.tex
\section{Indirect data delivery overview}
\label{sec:indirect-delivery}
           
In this section, we present the work that leverages the indirect data delivery approach. According to this approach, the data delivery results from the combined mobility of a sequence of non-controllable entities. The data moves along a route consisting of multiple segments, each traveled by a different entity. The works we review in this section are listed in the two last columns of Table~\ref{tab:classification}. We classify those approaches depending on the method they use to pass the data from one entity to another:

\begin{itemize}

    \item \textbf{Asynchronous method.} (Section~\ref{sec:indirect-async}) This method consists of passing the data indirectly via stationary or controllable mobile nodes. The stationary nodes act as data exchange relay points where entities can drop off data for later pick-ups by another entity. The movements of controllable nodes allow the data to be passed between two separated entities following non-intersecting routes.   
        
    \item \textbf{Synchronous method.} (Section~\ref{sec:indirect-sync}) 
    This method consists of passing the data directly from one entity to another while physically in contact. The objective is to decide whether the data should be passed or not every time two entities are in direct contact. As a result, the data can be passed to the first entity encountered, to a subsequent one, or to more than one encountered entities. The latter case results in data replications intended to improve the likelihood of delivery.

\end{itemize}

First, in Section~\ref{sec:indirect-async}, we review the asynchronous method useful for reliable transfers on entities with sparse movements and gaps in their connectivity. Second, in Section~\ref{sec:indirect-sync}, we review the synchronous method that leverages dense entity movements.

\section{Asynchronous method for indirect delivery}
\label{sec:indirect-async}

Instead of relying on the likelihood of two entities meeting together or on the utility of their contacts to move the data close to the destination, the asynchronous method consists of passing the data indirectly via intermediate nodes. We classify the approaches using this method depending on whether the intermediate nodes are stationary or mobile: 


\begin{itemize}
    
     \item \textbf{Stationary intermediate nodes.} A stationary intermediate node allows the data to be passed between two entities whose trajectories intersect without the entities being in contact at the same time. The intermediate nodes buffer the data so it can be passed asynchronously from one mobile entity to another. 
     
    \item \textbf{Mobile intermediate nodes.} In the case of entities following trajectories that intersect occasionally or never, the use of mobile intermediates nodes, such as message ferries or robots can bridge the gap between such entities. The route of a mobile intermediate node is calculated so the data can be passed asynchronously at various locations depending on the meeting points between the nodes and the entities.      

\end{itemize}

In the following, we present first the approaches relying on stationary intermediate nodes useful for dense numbers of nodes and second, those relying on mobile intermediate nodes useful for sparse networks.

\begin{figure}[h]
    \centering
    \includegraphics[width=0.85\columnwidth]{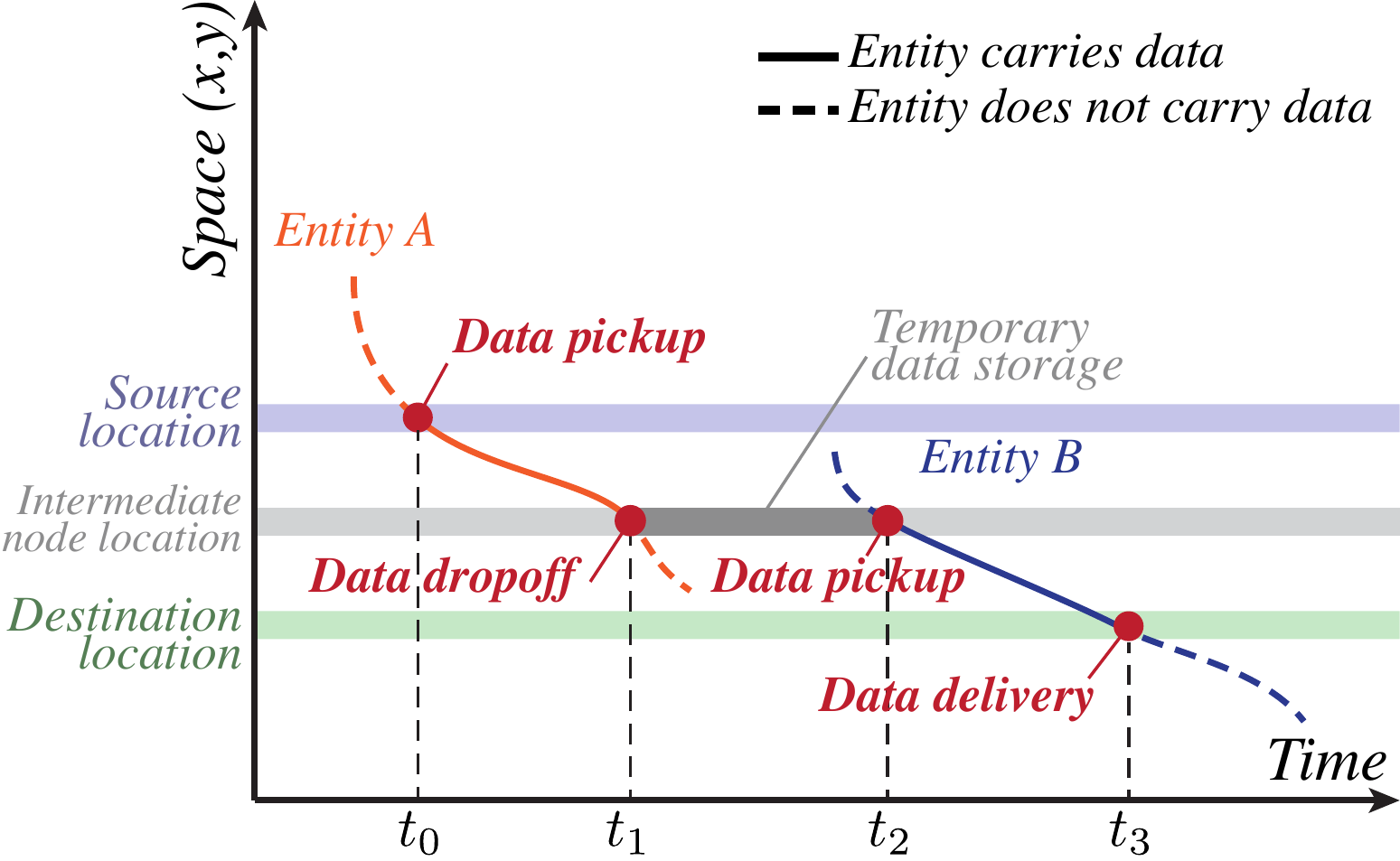}
    \caption{Asynchronous method indirect delivery using one intermediate stationary node located in between the source and destination locations. The intermediate node temporarily stores data dropped off by $A$ at time $t_1$ until $B$ picks it up at time $t_2$ to deliver it to the destination at time $t_3$.}
    \label{fig:dtn-indirect-async-geo}
\end{figure}

\subsection{Stationary intermediate nodes}
\label{sec:indirect-async-stationary}

The asynchronous method for indirect data delivery relies on the  trajectory composition of multiple entities without requiring the entities to be in contact. The data is passed asynchronously from one entity to another via intermediate stationary nodes commonly referred to as \textit{throwboxes}. A throwbox is equipped with wireless and storage capabilities enabling mobile entities to buffer the data hey carry before passed to another entity. In Figure~\ref{fig:dtn-indirect-async-geo}, we represent the asynchronous method with one stationary intermediate node located in between the source and destination locations. The trajectories of two entities $A$ and $B$ intersect with one another at the intermediate node location at times $t_1$ and $t_2$. $A$'s data dropoff and $B$'s data pickup results in the data passing from $A$ to $B$.

In the following, we first review the placement strategies proposed in the literature for improving the benefits of using throwboxes. 
We then present the research works that consider a collection of throwboxes already deployed without assuming any specific placement strategy. These works address the strategies 
for passing the data to the mobile entities in the transmission range of a throwbox. Those strategies may result in replicating the data if passed to multiple mobile entities. We present these strategies depending on the connectivity model considered between the throwboxes, as depicted in Figure~\ref{fig:throwboxes}.
We consider separately the case of the passing strategies relying on the prediction of the entity trajectories.    
Finally, we review the studies where disconnected throwboxes are public kiosks deployed for providing Internet connectivity to rural areas and inspect the approaches relying on intermediate nodes to offload large amount of data from the Internet for capacity improvement purposes.   




\subsubsection{\textbf{Stationary node placement}}
The placement of stationary intermediate nodes can improve performance metrics such as the delivery rate or delay. The following approaches study the deployment of intermediate nodes referred to as throwboxes or dead drops. Those nodes are supposed disconnected as depicted in Figure~\ref{fig:throwboxes-disconnected}.   

\smallskip\noindent\textit{Throwboxes.} In~\cite{zhao2006capacity}, the intermediate nodes are small inexpensive battery-powered devices called 
\textit{throwboxes}. Their objective is to improve both the transfer capacity between mobile entities and the data delivery delay. The authors formulate the throwboxes placement problem as a linear programming model based on the knowledge regarding the network structure including the traffic demands between each pair of entities and the complete list of contacts as well as the capacity resulting of those contacts, either between the entities or with the candidate locations for the throwboxes. They solve the placement problem by jointly calculating a multi-path solution for balancing the traffic load or the optimal logical route that offers the highest capacity for each traffic demand. Their simulation results using various mobility models for the entities, including synthetic models and one derived from real word mobility traces, show that the informed deployment of intermediate stationary nodes improves throughput and delay, especially when the entity movements are regular or when the data is dispersed on multiple paths.

\smallskip\noindent\textit{Dead drops.} Chawathe focuses on the data dissemination in sparse vehicular networks~\cite{chawathe2006inter}. He proposes to bridge the connectivity gaps between neighbor vehicles by the use of stand-alone devices, called \textit{dead drops}, located at road intersections. 
The author presents a greedy algorithm that selects a set of intersections where to deploy the dead drops by solving a minimum-weight $k$-set cover problem which minimizes the deployment cost of the dead drops while meeting the connectivity requirement among the flows of vehicles traveling different routes. 

\subsubsection{\textbf{Vector route calculation and data passing methods}}

\begin{figure}[t]

    \begin{subfigure}[b]{\columnwidth}
        \centering
        \includegraphics[scale=0.3]{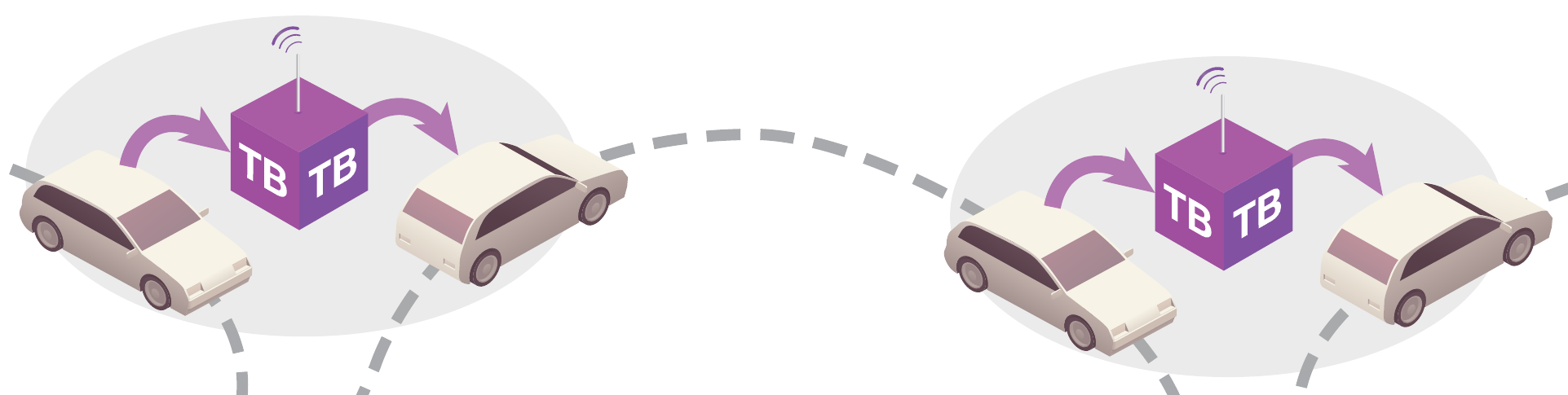}
        \caption{Disconnected throwboxes.}
        \label{fig:throwboxes-disconnected}
    \end{subfigure}
    
    \begin{subfigure}[b]{\columnwidth}
        \centering
        \includegraphics[scale=0.3]{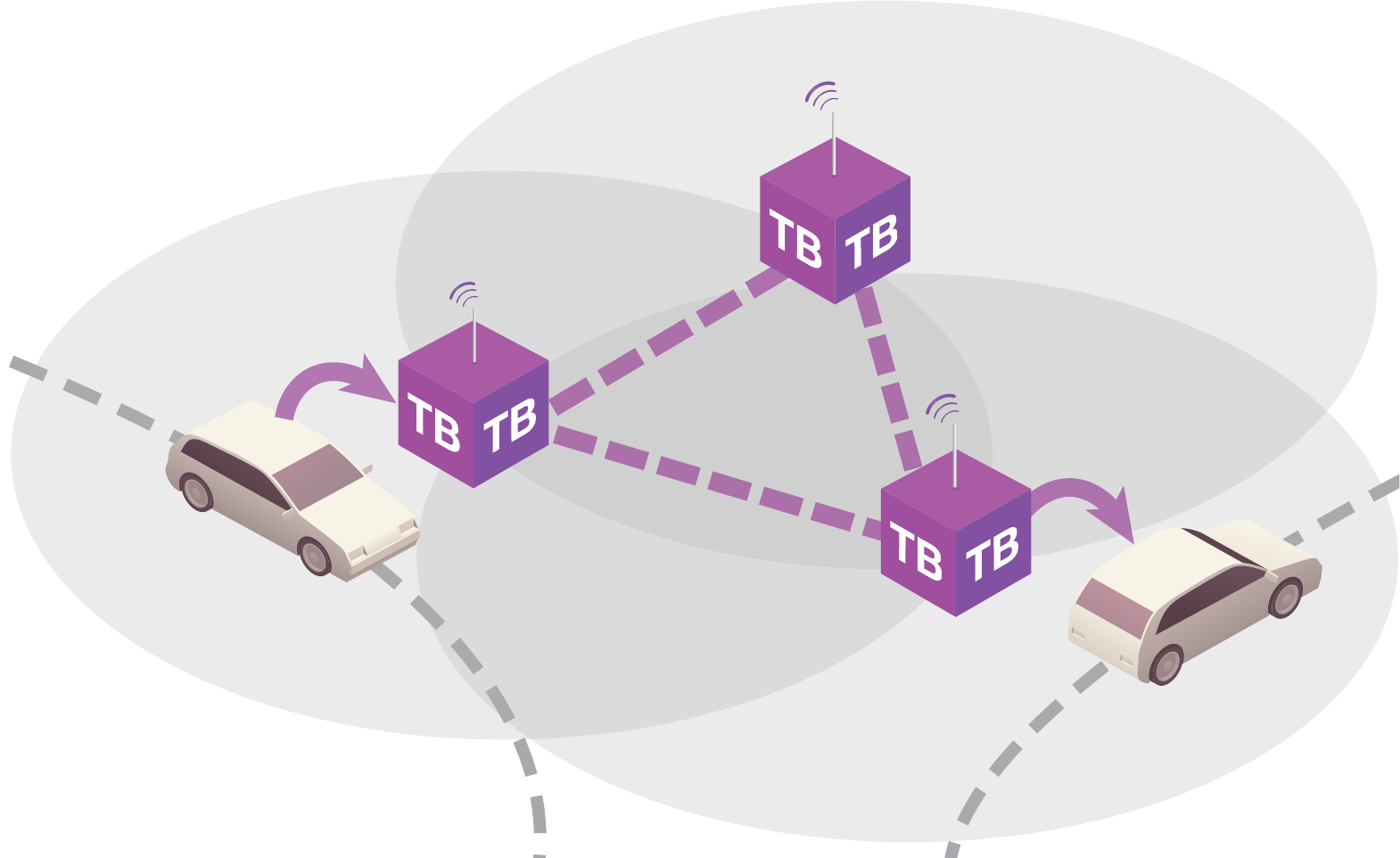}
        \caption{Mesh-connected throwboxes.}
        \label{fig:throwboxes-connected}
    \end{subfigure}
    
    \begin{subfigure}[b]{\columnwidth}
        \centering
        \includegraphics[scale=0.3]{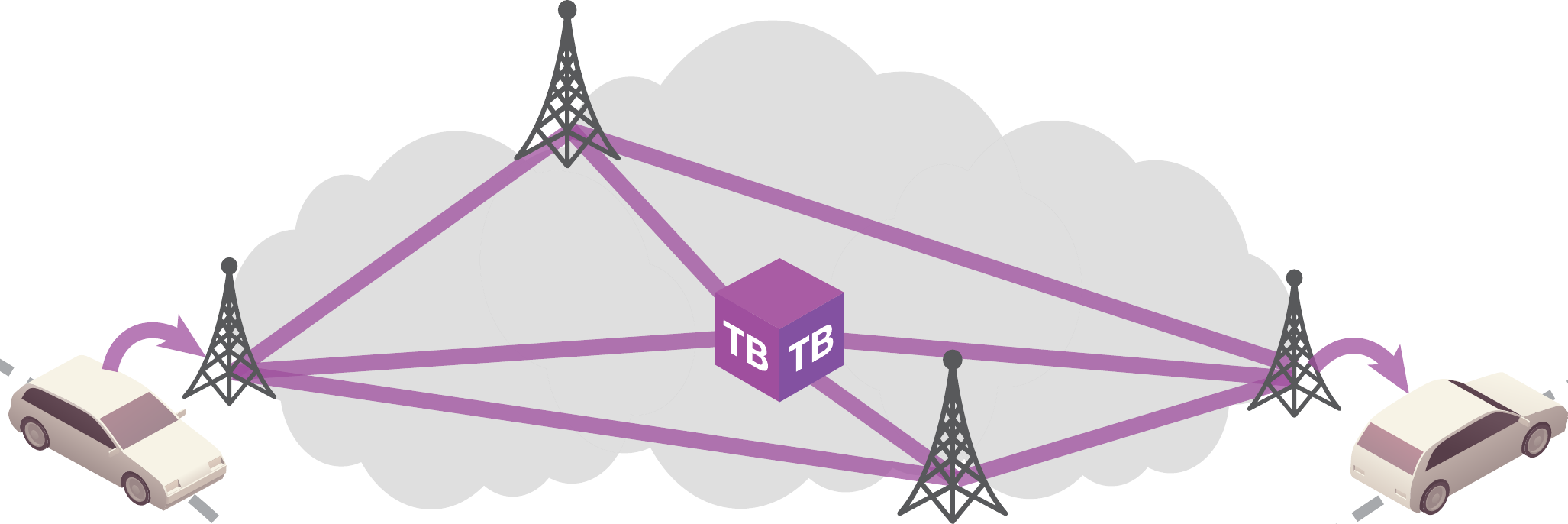}
        \caption{Single throwbox accessible by base stations.}
        \label{fig:throwboxes-base-stations}
    \end{subfigure}
    
    \caption{Throwboxes connectivity models.}
    \label{fig:throwboxes}
\end{figure}

Given an existing deployment of intermediate nodes, another aspect that has attracted research concerns the passing methods used to exchange data between a throwbox and the mobile entities in the throwbox transmission range. 
The following approaches have been proposed considering one or many throwboxes giving the various connectivity models depicted in Figure~\ref{fig:throwboxes}.  

\smallskip\noindent\textit{Disconnected pre-positioned throwboxes.} In the case of the disconnected throwboxes, the authors consider different routing strategies~\cite{zhao2006capacity}. Without any knowledge regarding the mobility of the entities and their contacts with the throwboxes, they propose an epidemic routing which consists of passing the data still stored at an intermediate node, to all subsequent visiting entities. They showed that the placement strategy used for deploying the throwboxes has little impact on their benefit in terms of transfer throughput with epidemic routing, given the number of data replicas generated which also incurs a poor utilization of the network resources. 

\smallskip\noindent\textit{Disconnected vs. connected randomly positioned throwboxes.} In~\cite{ibrahim2009analysis}, Ibrahim \etal consider the case of randomly placed throwboxes that can be either disconnected (Figure~\ref{fig:throwboxes-disconnected}) or fully connected (Figure~\ref{fig:throwboxes-connected}). In the latter, all throwboxes receive a copy of the data once stored at one location. Instead of calculating the vector routes followed by the entities, they evaluate multiple passing methods which allow the data to be passed from or to a mobile entity depending on who the entity is in contact with. These methods depends whether the contacts of the mobile entities happen with the source, the destination, the throwboxes, or other mobile entities. They found that passing the data between the mobile entities and the throwboxes improves the performance in terms of delivery delay and number of data replicas generated in the case of connected throwboxes, whereas passing the data between the mobile entities and the source or the destination performs better in the disconnected case.       

\smallskip\noindent\textit{Disconnected and connected throwboxes vs. base stations.} Banerjee~\etal compare the performance of the asynchronous method using various types of throwboxes 
to the synchronous method which consists of exchanging data between mobile entities in direct contact~\cite{banerjee2008relays}. They consider the cases of multiple throwboxes either disconnected (Figure~\ref{fig:throwboxes-disconnected}) or connected by a wireless mesh network (Figure~\ref{fig:throwboxes-connected}), and the case of a single throwbox where data is forwarded   using a conventional wired data network (Figure~\ref{fig:throwboxes-base-stations}). In the latter, the mobile entities offload their data via a set of base stations acting as gateways. 
In the disconnected case, the deployment of the throwboxes results in a placement strategy that depends on the density of mobile entities if known in advance. In the case of the wireless mesh network, the connected throwboxes are placed within range of one another. 

The authors evaluate the performance of various passing methods between the mobile entities and the intermediate nodes. They consider two passing methods: (\textit{i}) a two-hop method, which requires the data to be passed from one mobile entity to another via a stationary intermediate node, and (\textit{ii}) a random epidemic method, where mobile entities or intermediate nodes pass the data to randomly selected subsequent mobile entities. These methods are compared to the RAPID routing protocol~\cite{balasubramanian2007dtn}; we will present more details in Section~\ref{sec:indirect-sync-non-anchored}. The decision to pass data is taken in order to optimize a routing metric such as the delivery delay or ratio. This decision is based on the distributed knowledge regarding the current number and location of the data copies resulting from previous passing. The authors found that the addition of base stations improves the performance of the system and reduces data delivery delay by a factor of two. In order to have similar performance improvements, the system must rely on twice as more throwboxes connected by a mesh network or five times more disconnected throwboxes. They also found that few stationary intermediate nodes bring more benefits compared to the large number of mobile entities otherwise needed to transport the data. 

\subsubsection{\textbf{Trajectory prediction}}
In the following work, the intermediate stationary nodes evaluate the decision of passing data to a visiting entity by predicting the trajectory of the entity. 

\smallskip\noindent\textit{Distance-vector routing with landmarks.} With DTN-FLOW~\cite{chen2015dtn}, the authors propose to improve the carrying utility of mobile entities by using stationary intermediate nodes they call \textit{landmarks}, placed in well visited areas. The objective is to capitalize on the frequent visits of mobile entities who have a higher probability of reaching an intermediate landmark than the final destination. As a result, the data is carried hop-by-hop through a sequence of landmarks by different entities before reaching its final destination. The design of DTN-FLOW consists of placing the landmarks. The routes offering the shortest delay across the network of landmarks are then calculated. The authors use a distance-vector approach relying on the mobile entities to measure the transit delay and carry the distance vectors between neighbor landmarks. Finally, the resulting routing tables are used in combination with mobility predictions based on the entity visit history to forward the data across the network of landmarks toward their final destination. 

\smallskip\noindent\textit{Energy-efficient data passing.}
Banerjee \etal study the problem of power management to increase the lifetime of throwboxes powered by solar-charged batteries in the context of a vehicular network~\cite{banerjee2007energy}. They propose a prediction model which measures the utility of contacts with respect to the data delivery rate and latency. This model requires the mobile entities to beacon periodically their position, direction, and speed using long-distance radio. These periodic beacons are used by a throwbox to predict the trajectory of incoming entities. They then introduce a scheduler which selects a subset of all contacts so as to minimize the number of data replications and thus the throwboxes' energy consumption while maximizing the number of successful deliveries.

\subsubsection{\textbf{Routing with public kiosks in rural areas}}

\begin{figure}
    \centering
    \includegraphics[width=0.8\columnwidth]{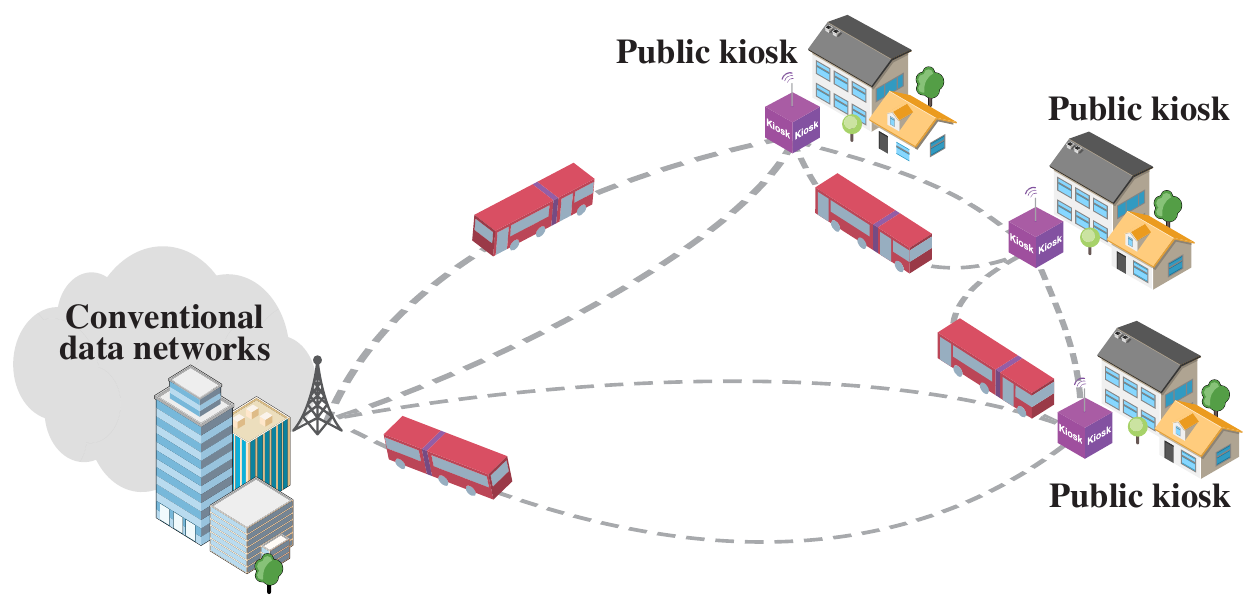}
    \caption{Internet access via public kiosks for rural areas.}
    \label{fig:public-kiosks}
\end{figure}

The concept of public kiosks depicted in Figure~\ref{fig:public-kiosks} allows users in rural areas to access the Internet by dropping off their data. The data is then brought by buses to the closest city with access to the Internet. Contrary to the approaches we presented in Section~\ref{sec:direct-passive-scheduled} such as Daknet, here the data can  be passed via multiple intermediate nodes, including the public kiosks and the Internet access points. 

\smallskip\noindent\textit{Flooding and reverse-path forwarding.} 
In KioskNet~\cite{seth2006low}, the authors propose different routing strategies so a user can receive the data they requested back from the Internet. Those strategies include flooding which guarantees user reachability but does not scale, and reverse-path forwarding which requires users to send register messages toward the closest Internet access point. 
The register messages are carried by buses along a default route connecting a sequence of stationary nodes including public kiosks and the Internet access points. Each time an intermediate node receives a register message, a state pointing toward the previous node where the register message was buffered is installed. 
These states are used to forward the data back to the user along the default route in reverse. 
The authors also consider the use of link state routing to avoid relying on default routes and for better resistance to link failures. The expected delay between two adjacent intermediate nodes serves as the link weight and is flooded in link state packets.

\smallskip\noindent\textit{Link state routing.} Following a similar architecture with KioskNet, Demmer and Fall propose DTLSR~\cite{demmer2007dtlsr}, a link state routing protocol which exploits the regularity of the contacts between buses and the kiosks or the Internet access points. From their point of view, the predictability resulting from buses' scheduled mobility implies a regular stable topological structure which allows the use of a link state-based route calculation. 
They propose a modified link state protocol which draws on the buffering capabilities of the nodes to pass the data once a link is available again. Various link metrics, including bandwidth, delay, and queue are flooded with the link state packets. Nodes maintain the topology of the network including the link metrics and the neighboring relationships used by Dijkstra shortest path calculation. Instead of considering a non-answering neighbor node permanently unreachable, the LSA lifetime is increased to last at most one year, allowing the path calculation to account for the links dynamic history and to consider a broken path as valid even if not available at some point of time.     

\smallskip\noindent\textit{Postal data mailings.}
Already introduced in Section~\ref{sec:internet-improv}, Postmanet considers multiple routing strategies using an indirect delivery model between groups of sending and receiving kiosks~\cite{wang2004turning}. A first strategy involves the use of an intermediate distribution server in charge of maximizing the utilization of the daily mail delivery and collection rounds. Data can be either aggregated from or replicated to multiple memory devices depending on the number of senders or receivers. In a second strategy, they consider the use of multiple data distribution centers that communicate together to take coordinated routing decisions. 


\subsubsection{\textbf{Enhancing Internet capacity using traffic offloading}}
The principle of traffic offloading was already introduced in Section~\ref{sec:contracted-mobility} where we presented approaches using the services of a package delivery company to ship data between the source and destination. In the following, we review the approaches which rely on intermediate stationary nodes, such as train stations. The intermediate stationary nodes intersperse a sequence of logical links resulting from the mobility of trains in the first work we present and of courier deliveries in a second work. 

\smallskip\noindent\textit{Intermediate node fair resource allocation.} In TrainNet~\cite{zarafshan2010trainnet}, the authors propose to use the railway lines as an offloading channel. Trains are equipped with hard disks and carry large amounts of data across a sequence of train stations before reaching their final destination. The train stations are also equipped with hard disks where data can be stored before transferred to the next train operating on another line. Given the limited capacity of the train and station storage, they formulate the routing problem of carrying data on trains as a fair resource allocation problem. This formulation allows the maximization of the data throughput and hard disk utilization, while minimizing the delay and data losses.

\smallskip\noindent\textit{Budget-constrained traffic offloading using courier deliveries.} Cho and Gupta propose Pandora, an intermodal data delivery system which uses the services of a courier company service in tandem with the Internet so as to minimize the cost and latency of bulk data transfers~\cite{cho2011budget}. They consider that a transfer follows a sequence of Internet links interspersed by courier deliveries. They account for the holdover time needed to transship the data from one means of transport to another which includes transferring the data on disks and packaging the disks for Internet-courier transshipments and the inverse procedure for courier-Internet transshipments. Their objective is to solve a budget-constrained transfer problem that minimizes the transfer duration subject to a budget constraint. They represent the sites connected by both shipping links and Internet links as a dynamic flow graph. The links are characterized by their capacity, cost, and transit time, which vary over time in the case of shipping links (\eg both the cost and transit time vary as a function of the level of service~---~Overnight, Two-day, Ground). The authors model the planning problem as a flow over time problem~\cite{fleischer2007quickest}. They solve this problem using optimization designed for large-scale networks. Their results show that Pandora provides better performance compared to transfers using only one means of transportation, whether this means is the Internet or the courier service alone. 


\subsection{Mobile intermediate nodes}
\label{sec:indirect-async-mobile}


\begin{figure}[t]
    \centering
    \includegraphics[width=0.85\columnwidth]{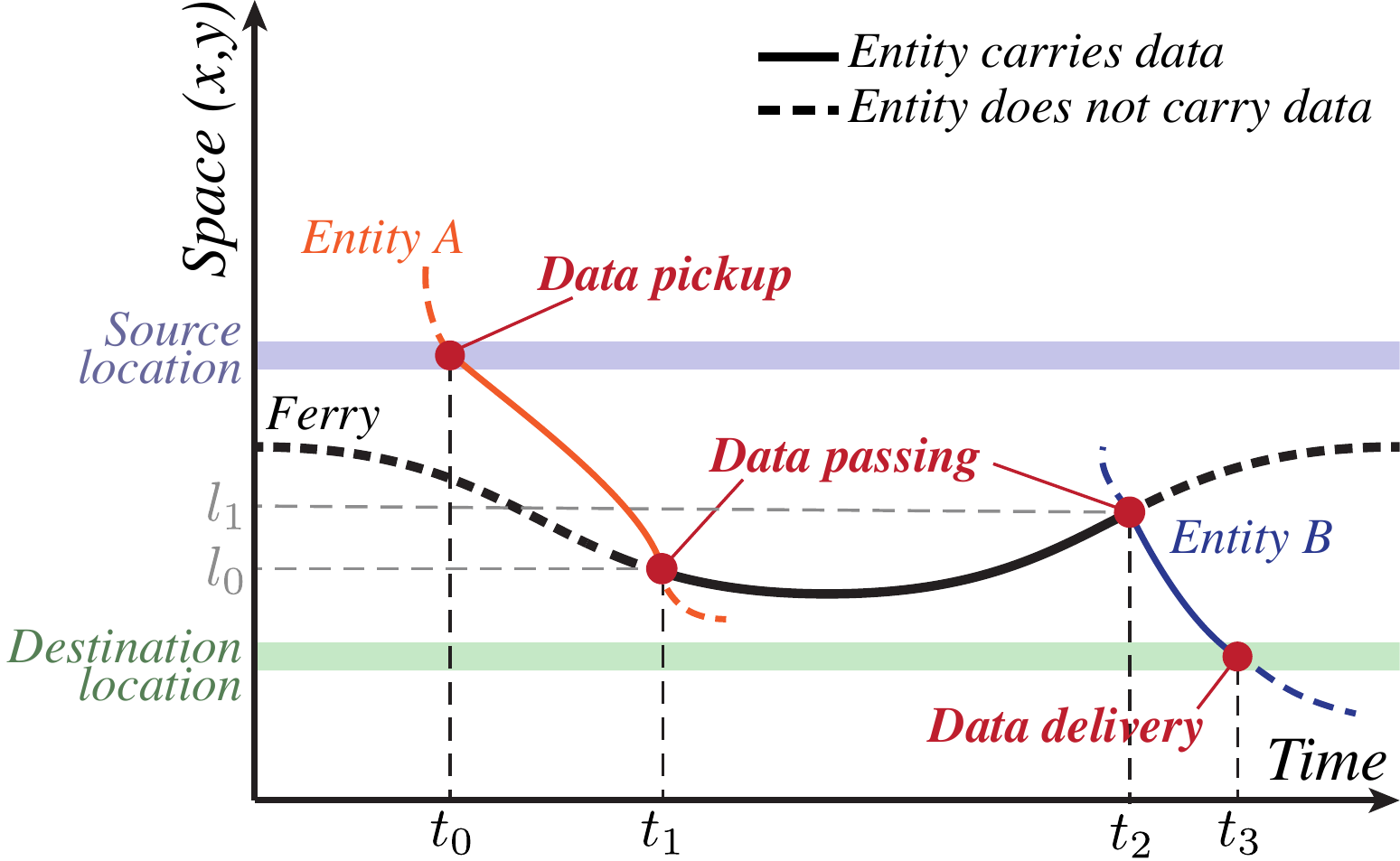}
    \caption{Asynchronous indirect delivery between the two entities $A$ and $B$ using an \textit{intermediate mobile node} (\ie a ferry). The ferry temporarily stores the data passed by $A$ until it comes in contact with $B$ and passes the data before its final delivery. In this case, the data passing happens at any location, as long as it is on the trajectory of the ferry.}
    \label{fig:dtn-indirect-async-nongeo}
\end{figure}
In this section, we review the approaches which rely on mobile intermediate nodes 
in charge of bringing the data after meeting with a first entity to a second one. When the intermediate nodes move along a pre-calculated trajectory they are referred to as \textit{data ferries}. They are a special case of controllable entities already introduced in Section~\ref{sec:direct-control}. While a controllable entity delivers data end-to-end (\eg delivers the data collected from one or many sources to a destination such as a sink), mobile intermediate nodes carry the data on behalf of two mobile entities acting as the source and destination of the data. 

In Figure~\ref{fig:dtn-indirect-async-nongeo}, a ferry comes in contact with entity $A$ at time $t_1$ and location $l_0$ before moving towards entity $B$ and comes in contact at time $t_2$ and location $l_1$. As a result, the data is passed asynchronously between $A$ and $B$ at two different points of time, $t_1$ and $t_2$. Mobile intermediate nodes enable remote data passing between two entities with non-intersecting trajectories. 


\subsubsection{\textbf{Scheduling predefined contacts between a single data ferry and mobile entities}}
The approaches relying on a single ferry can be classified depending on which one of the entities or the ferry changes its course to meet with the other. 

\smallskip\noindent\textit{Mobile entity-initiated encounters.} In~\cite{zhao2004message}, Zhao \etal exploit the benefit of using a single data ferry to enable communication opportunities with mobile entities. They study two different ferrying schemes including NIMF (Node-Initiated Message Ferrying). 

In NIMF, the mobile entities move in order to get close and create communication opportunities with the ferry who keeps the course of its trajectory. The mobile entities meet a ferry along the shortest path they follow as a result of a decision function which minimizes the message drops and the negative impact of detouring from their route. Entities are assumed to be mobile for accomplishing assigned tasks in the deployment area, under the constraint of limited resources such as battery, memory, and computation power.



\smallskip\noindent\textit{Ferry-initiated encounters.} Instead of modifying the course of the mobile entities, the following ferrying schemes require the ferry either to adjust its trajectory or to make stops so the mobile entities can meet.   

In~\cite{zhao2004message}, Zhao \etal present FIMF (Ferry-initiated Message Ferrying) where a ferry with no buffer or energy constraints beacons its current position while moving along a pre-defined default route. The periodic beacons are used by the mobile entities 
to determine if the ferry is close. A mobile entity 
transmits a series of meeting requests containing its current position so the ferry can adjust its trajectory and meet with the entity. Both beacons and requests are transmitted via a long range radio.  
The authors model the ferry route problem by adapting the Minimum Latency Problem to define in which sequence the ferry visits the requesting entities so as to minimize the message drops resulting from message timeouts or buffer overflows. To solve this NP-hard problem, they propose two heuristics. The first consists in selecting the nearest entity as the next one to visit to minimize message drops. The second uses local optimization techniques to adapt the Travel Salesman Problem (TSP) algorithms used in~\cite{zhao2003message}.

Their simulation results show that both NIMF and FIMF ferrying schemes perform better than epidemic routing with regard to data delivery and energy consumption. 

In~\cite{bin2006message}, Bin \etal exploits the regular route taken by a ferry along which the ferry makes stops at predetermined locations called \textit{way-points}. While moving or waiting at the way-points, the ferry makes contact with mobile entities that have messages to send or receive. The authors first identify the locations for the way-points that minimize the time the ferry spends waiting while maximizing the probability of meeting mobile entities. This requires knowledge regarding the structure of the entity mobility. They then calculate the route connecting the way-points by solving the Travel Salesman Problem (TSP) with two heuristics depending on the distribution of the mobile entities across the area. The first favors the shorter route when entities are uniformly spread in the deployment area. The second favors way-points closer to the center of the area when entities are not uniformly spread across this area.


\subsubsection{\textbf{Controlling multiple ferries}}
The three previous approaches consider the tours made by a single ferry. Adjusting the route of multiple ferries in order to meet with mobile entities creates additional challenges. The scheduling of these meetings can be reduced to a dial-a-ride problem~\cite{savelsbergh1985local}. 

In~\cite{burns2008mora}, Burns~\etal borrow two multi-objective control techniques from robotics to calculate the near-optimal routes of ferries they call \textit{agents}. They first calculate the routes by optimizing an individual performance metric such as increased bandwidth or reduced delay. These routes are then combined using the multi-objective control techniques which balance the individual optimization objective of each route concurrently. They consider the subsumption composition, which prioritizes the ordered metrics and optimizes them successively, starting from the higher ones, and each up to a given performance threshold. They also consider the nullspace composition, which outperforms the subsumption composition and orders the metrics such that the optimization of these lower metrics does not affect the performance of the higher metrics. 

\subsubsection{\textbf{Maintaining a global view of the network state}}
To decide in which order to visit the entities, the ferries need an up-to-date view of the network state which includes the position of the entities or the position of the ferries. 

\smallskip\noindent\textit{Single ferry.} In~\cite{zhao2004message}, this view includes the location of the entities and the ferries as well as the message generation and drop rates. In FIMF (Ferry-initiated Message Ferrying), the ferry periodically announces its location using long range radio. Close-by mobile entities also use long range radio to schedule a meeting by sending meeting requests including their updated positions until the ferry comes meet with them. In that regard, FIMF depends on the accuracy of  the information about the ferry position. The meeting requests also contain the local message drop rate for the sending entities. In NIMF (Node-initiated Message Ferry), the mobile entities are assumed to know the route followed by the ferry in advance and decide to move proactively in order to meet with the ferry. The route can be either broadcast by the ferry using long range radio or learnt via other out-of-band means. In NIMF, the ferry also broadcasts the rates at which messages are generated or dropped for each destination mobile entity.    


\smallskip\noindent\textit{Multiple ferries.} With multiple ferries, Burns \etal rely on a distributed approach so the ferries can estimate a global state information about all network participants including the other ferries and the mobile entities~\cite{burns2008mora}. The mobile entities maintain information about their own state. Ferries update their view by collecting such information each time they meet with a mobile entity. Ferries also exchange information about all participants when they meet with other ferries who may have a more up-to-date view. This view includes information about the messages carried on behalf of each participant and the location and time stamp of the last encounter. Though this approach does not require prior knowledge regarding the ferries' movements or the structure of the entity mobility, it requires all participants to have synchronized clocks and depends, as for FIMF, on the accuracy of locating the positions of the entities.  


\subsubsection{\textbf{Non-controllable dedicated entities}}
While the ferries or robots are dynamically controlled to decide which entity to visit, other approaches propose strategies to use the already-existing movements of dedicated entities that act as ferries to transport the data. 

In the context of Micro Aerial Vehicles (MAVs) used for Search and Rescue missions, Asadpour \etal propose a forwarding algorithm to route the data recorded by hovering MAVs equipped with onboard cameras to a central ground station~\cite{asadpour2016route}. The authors combine the use of relay MAVs which are connected via traditional wireless links and ferry MAVs which carry the data closer to the destination. The forwarding algorithm they propose relies on linear short-time prediction using the MAVs' position, speed, and direction information shared on a long range, but low throughput out-of-band radio channel. 

They first estimate the throughput of both the physical links between two relay MAVs within transmission range and the throughput of the virtual links resulting from a ferry MAV that carries the data to the next MAV or the destination. The algorithm then forwards the data along the multi-hop shortest path to the destination, if any. Otherwise, a greedy geographic forwarding is used so the data is carried to the destination or to a next MAV closer to the destination. While in both cases, the forwarding decisions are only based on current locations, they also propose two heuristics based on the future locations of MAVs. The first heuristic estimates for all MAVs their future proximity to the destination, while the second estimates the capacity of a link by predicting its connection time.  



%% file: 3c-indirect-sync.tex
\section{Synchronous method for indirect delivery}
\label{sec:indirect-sync}

In this section, we review the synchronous indirect delivery model which requires the mobile entities to be physically in contact at the same time for the data to be passed. This model captures the strategies introduced for Delay-Tolerant Networks (DTNs) and other close variants such as Opportunistic Mobile Networks (OMN). The data follows a vector route consisting of segments of trajectories each followed by a sequence of different mobile entities. The data is passed between two consecutive entities when they meet, where their trajectories intersect. Given the large amount of literature available in the wide area of DTNs, we focus on the relevant approaches with regard to our survey.
  
We classify the synchronous indirect delivery approaches depending on whether entities pass the data when they meet at any location or when they meet at pre-defined locations:


\begin{itemize}
    \item \textbf{Floating composition.} In the floating case, the data can be passed anywhere as long as the mobile entities are in direct contact. On a contact, the different strategies help determine whether to pass the data, keep it, or replicate it to the other node.
        
    
    \item \textbf{Pre-positioned composition.} In the location-dependent case, the decision to pass the data is taken when mobile entities meet at specific locations. Those locations are determined given some specific properties such as the contact density.
    

\end{itemize}

In the following, we first present the strategies used to pass the data in the floating case relevant for networks with unknown node movement patterns and then, in the pre-positioned case for networks with known node movement patterns. 



\subsection{Floating composition}
\label{sec:indirect-sync-non-anchored}

\begin{figure}[t]
    \centering
    \includegraphics[width=0.85\columnwidth]{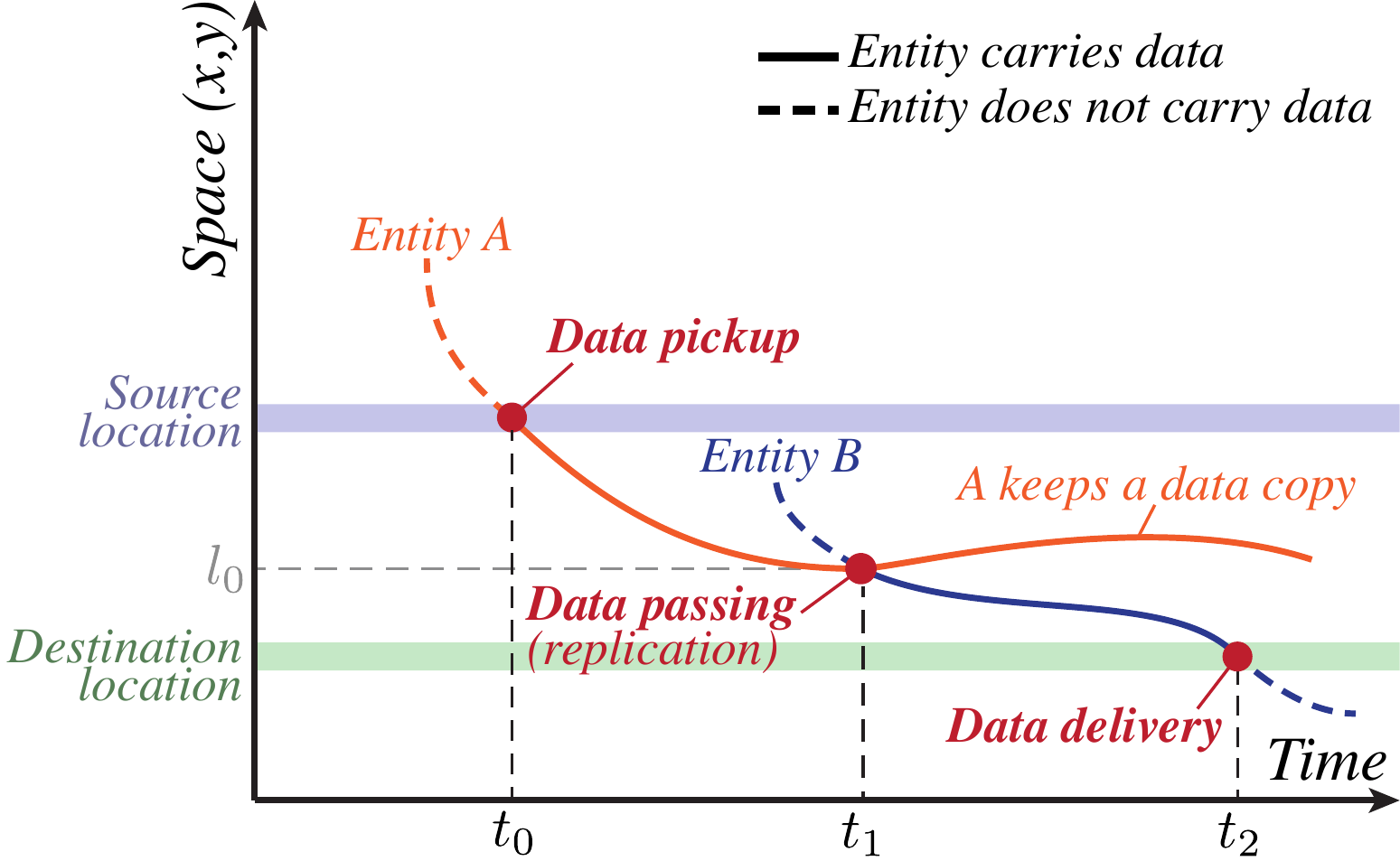}
    \caption{Indirect synchronous floating composition between two entities $A$ and $B$. The data passing happens at any location, as long as the entities come in contact.}
    \label{fig:dtn-indirect-sync-nongeo}
\end{figure}

When two entities are in direct contact, this creates an \textit{opportunity} to pass the data which may help bring the data closer to the destination. We represent the floating case in Figure~\ref{fig:dtn-indirect-sync-nongeo} where two entities $A$ and $B$ come in contact at location $l_0$ and time $t_0$. $A$ picked up the data at the source and decided to pass it to $B$ when they came in contact. $A$ could have also decided to discard the contact opportunity with $B$ and postpone the passing to another entity. In the depicted scenario, $A$ duplicates the data and passes a copy of the data to $A$ at time $t_0$. This results in replicating the data. The data thus takes a multipath route consisting of the different paths followed by the copies. In our figure, the data follows a first segment which matches the trajectory of $A$ before reaching location $l_0$ where the replication occurs as $A$ comes in contact with $B$. The original copy follows the route that corresponds to the trajectory followed by $A$, while the new copy follows the route taken by $B$ after $t_0$. These multipath routes are the result of the synchronous composition of $A$ and $B$'s trajectories. 


In this section, we present different classes of passing strategies according to whether they require a control plane and whether the control plane is local or shared with other entities. The control plane relies on the level of knowledge of the state of the entities, such as the number of copies of the data in transit or the knowledge of the entity mobility structure.

Similar instances of these classes were studied by Jain \etal~\cite{jain2004routing} with the objective of comparing the  routing performance in a delay tolerant network under various assumptions with respect to the knowledge required regarding contacts, queue occupancy, and traffic demands. 
They consider three classes including the zero knowledge and the complete knowledge classes. The latter class refers to the complete knowledge of the future encounters between entities. Since this class is not achievable in practice, the authors propose the partial knowledge class with weaker and more realistic assumptions regarding the required level of knowledge.  
Within each of the three classes, the authors propose various passing strategies they compare in terms of delivery delay and delivery ratio. 


In the following, we propose a classification of the passing strategies depending on whether they require a control plane and if so, whether the control plane is \textit{local} to each entity or \textit{shared} among the entities.


\subsubsection{\textbf{No control plane}}
Without any control plane, the strategies consist in passing the data whenever two entities meet together without any knowledge regarding the state or the mobility structure of the entities. The lack of control plane prevents informed passing decisions and is related to the nature of the mobile entities. The approaches using this strategy exploit the existing movements of non-controllable entities whose mobility is used passively. 

\smallskip\noindent\textit{First contact.} Jain \etal propose the first contact strategy where the data is passed to the first entity encountered~\cite{jain2004routing}. The data is removed from the buffers of the passing entity so a single copy of the data is in transit. The choice of the next-hop entity being ``random'', the delivery of the data is uncertain as the data may not make any progress toward the destination.



\smallskip\noindent\textit{Epidemic routing.} At the other extreme of the spectrum is the epidemic ``routing''. Although referred to as a routing scheme, it relies on a blind passing strategy~\cite{vahdat2000epidemic,papdopouli2000seven}. Every relay entity that receives a copy of the data will further replicate it by passing a new copy to all subsequent entities they meet with and so on. The data is replicated and spread among multiple paths, each followed by a replica that may reach the destination. If among all these paths, there are some that lead to the destination, the shortest will guarantee the minimum delivery delay. The resulting replications are thus intended to alleviate the lack of knowledge regarding the mobility of the entities assumed to move randomly.
However, under limited resources, including buffer storage, epidemic routing is not efficient as it creates an exponential number of replicas that fill the buffers of the entities. 7DS, an ad-hoc peer-to-peer data sharing system, also relies on epidemic routing to share the data queries or the data itself among the mobile nodes (or peers)~\cite{papadopouli2001design,papdopouli2000seven}. With the power conservation mode at the mobile nodes, the authors propose an improvement to the routing scheme by introducing a short delay of a few seconds before broadcasting a message again.

\smallskip\noindent\textit{Data MULEs and Infostations.} Epidemic routing is one of the indirect passing strategies proposed in the context of mobile sensor networks, as well as with the Data MULE (Mobile Ubiquitous LAN Extension) architecture already presented in Section~\ref{sec:direct-passive-random}~\cite{shah2003data}. According to this strategy, a mule can pass the data it carries to another mule whenever they meet together. 


The Infostation model presents a data MULE application where terminals carried by independent mobile users connect to Infostations acting as access points distributed over a geographical area~\cite{goodman1997infostations}. When in the vicinity of an Infostation, the mobile terminals transmit at very high rates, thus trading coverage for capacity. 

With SWIM (Shared wireless Infostation model), Small and Haas enhance the Infostation model with indirect delivery~\cite{small2003shared}. Radio-tagged whales collect biological and environmental data continuously. The collected data is transmitted when the whales are in transmission range of buoys acting as Infostations located in known feeding grounds. Since the typical dive times of a whale may vary from few minutes to hours, whales are equipped with large amount of memory where data is stored until offloaded to Infostations at higher data rates. To improve the delivery delay of the collected data, copies of the data are passed from one whale to another when they are grouped together.  
To avoid excessive data replication, the data is passed in a probabilistic manner with a fine-tuned probability which guarantees that the data will be transmitted to an Infostation within a target duration since its generation.

\smallskip\noindent\textit{Vehicular sensing systems.} CarTel~\cite{hull2006cartel} proposes CafNet, an indirect delivery mechanism for traffic and environmental monitoring. CarTel takes advantage of the trips taken routinely by private cars to turn them in data mules. Equipping cars with sensors provides a cost-effective way to collect environmental data or to record traffic information. Encounters between cars are opportunities to exchange the data they have each collected as one may have a higher likelihood of passing by a gateway where the data should be uploaded. BikeNet proposes a similar approach where data mules are bikes~\cite{eisenman2009bikenet}. Note that in both works, no passing strategy is implemented as they consider a direct delivery approach in their evaluations.

\subsubsection{\textbf{Local control plane}}
The following approaches rely on a control plane local to each entity to improve the performance of epidemic routing by limiting the excessive usage of the network resources. Each entity acquires information regarding other entities, the destination, or the network in its whole. With a local control plane, the knowledge acquired by each entity results in their local view. 


\smallskip\noindent\textit{Encounter history-based routing.} In the following approaches, the passing strategies rely on mobility statistics gathered by each entity regarding previous encounters or visited locations.  

In~\cite{spyropoulos2005spray}, Spyropoulos \etal propose the Spray-and-Wait protocol to bound the number of replications using two phases~\cite{spyropoulos2005spray}. In the ``spray'' phase, a given number of copies are transmitted from the source to other entities acting as relays who can in turn transmit more copies. In the ``wait'' phase, the relays wait to encounter the destination to pass the data. They show that the number of copies required to achieve an expected delivery time depends on the number of mobile entities in the network, as more copies will be needed for denser networks. For the entities to be able to estimate this number, the authors propose to equip them with a local control plane consisting of a method compatible with mobility models with exponentially distributed meeting times, such as random walk. For each entity, this method consists  in estimating the average inter-meeting time with the other entities. To this end, all entities keep a record of the other entities they have encountered.

The Spray-and-Focus protocol~\cite{spyropoulos2007spray} uses the same ``spray'' phase as in the Spray and Wait protocol~\cite{spyropoulos2005spray}, followed by a ``focus'' phase where the relay entities can forward a copy of the data to other entities before meeting the destination. The decision to forward a copy of the data in the focus phase is based on a utility function which measures the usefulness of an entity in getting the data closer to its destination. For each entity, this utility function maintains the list of entities they encountered and the time since their last encounter. With the meets and visits (MV) protocol~\cite{burns2005mv} and the drop-least-encountered (DLE) protocol~\cite{davis2001wearable}, forwarding decisions are taken depending on the meeting frequency with other entities. To this end, each entity keeps track of its previous encounters with other entities. The MV protocol further increases the likelihood of delivering the data through a path of meetings by considering their meeting frequency but also with what frequency they visit specific regions. To this end, the entities keep track of their visits to regions defined by a cell-based geographic grid.

ZebraNet is a platform to monitor zebra wildlife in Kenya and track their location history logged in collars~\cite{juang2002energy}. The data logged on the collars is collected when the zebras are in the transmission range of base stations, which are either fixed or mobile. In the latter case, the base stations are carried by researchers who periodically drive-by or fly-by to collect the data from the animals. To reduce energy consumption, the collars alternate between standby and active modes. When active, the collars scan their surroundings in search of other zebras or base stations. To decide to which zebras to pass the data, the authors propose a history-based protocol which ranks the zebras according to previous encounters with the base stations. Each zebra ranks itself according to the number of consecutive scans that found a base station. Conversely, this rank is decremented in case a scan fails in finding a base station. The data is passed to the zebra with the highest score to increase the chances of successful delivery. ZebraNet also uses delete-lists to remove the remaining data that has been successfully delivered. These lists are updated with a gossip protocol whenever two zebras are in contact. They allow discarding copies of data that were successfully delivered to free up the buffers and avoid overflows.

\smallskip\noindent\textit{Geographical routing.} Other protocols such as GeOpps~\cite{leontiadis2007geopps} and GeoSpray~\cite{soares2014geospray} exploit the geographical information provided by the navigation system of vehicles to forward the data towards its destination and minimize the data delivery time. A vehicle broadcasts the destination of the data it carries to its neighboring vehicles. The data is passed to another vehicle following a shorter route to the destination or a route passing closer to the destination. As a result, the data follows a vector path which consists of adjacent segments of trips each taken by vehicles whose trajectories intersect. The data is passed from one vehicle to another when they meet at those intersection points. Note that vehicles are assumed to follow the route suggested by their navigation system.    

\begin{figure}
    \centering
    \includegraphics[width=0.9\columnwidth]{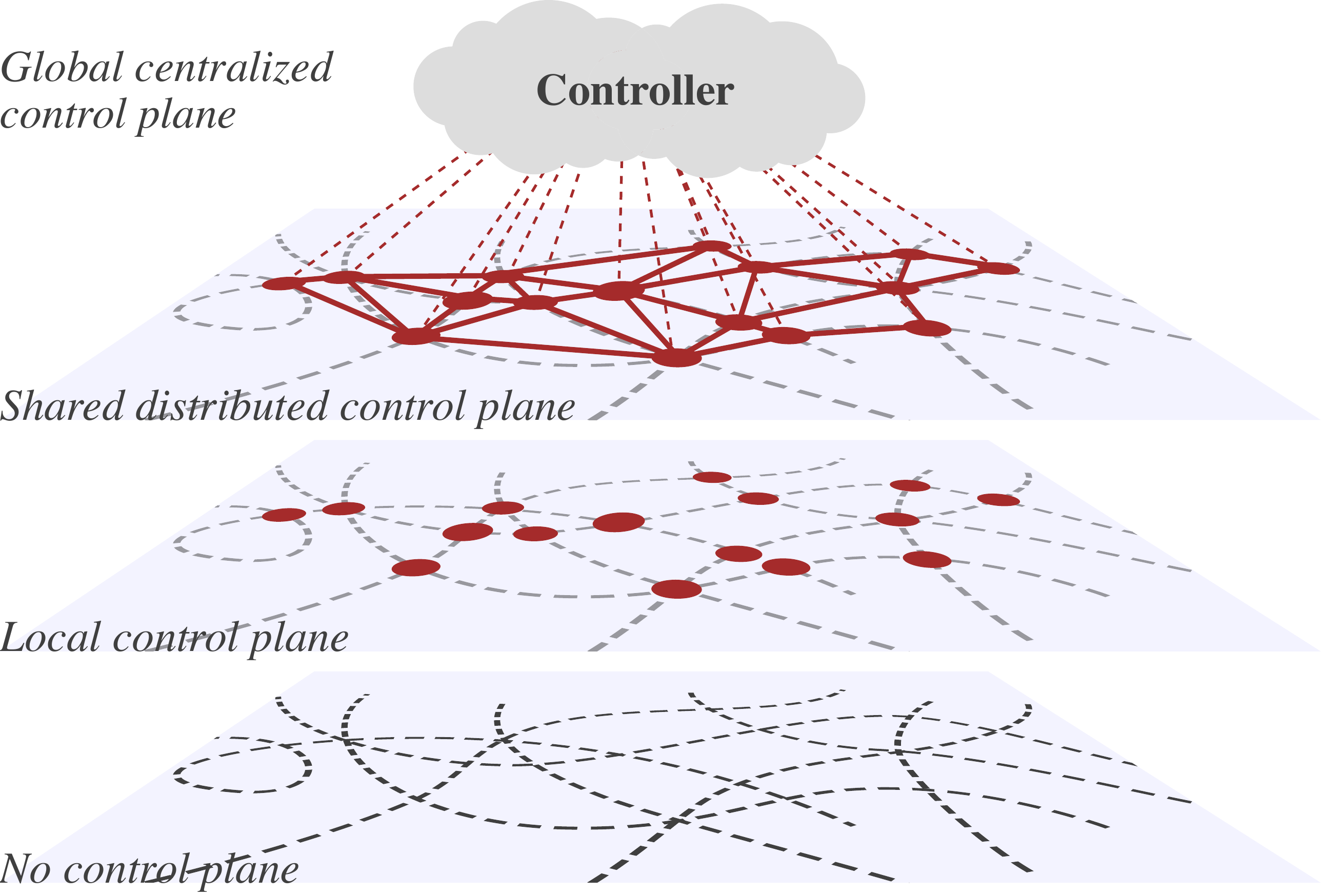}
    \caption{Different levels of control plane.}
    \label{fig:control-planes}
\end{figure}

\smallskip\noindent\textit{Social-based routing.} BUBBLE is a data passing algorithm designed for networks where entities are human-carried devices such as smartphones~\cite{hui2011bubble}. The movements of the entities are thus dictated by their users. BUBBLE proposes to measure the ability of a mobile entity to relay the data toward the destination by considering the social interactions of its user. Entities are ranked according to two social metrics, namely the communities to which a user belongs and the centrality which captures its tendency of meeting with more people. BUBBLE selects a relay entity if ranked with a higher centrality than the current carrying entity. Once the data is in the hands of an entity in the same community with the destination, the data is passed to other entities of this community. The authors propose distributed methods for each entity to be able to detect the communities it belongs to and to calculate its centrality. Note that the authors also proposed to use centralized community detection methods that are more efficient, but difficult to deploy, as they require a global state of the system shared by all the nodes. 

\subsubsection{\textbf{Shared control plane}}
With a shared control plane, the passing strategies require a knowledge that exceeds the local view of each entity. Entities collaborate by exchanging the information they acquired so to create this knowledge. Note that the resulting view varies from one entity to another depending on their encounter history.   


In PRoPHET, the mobile entities implement a shared control plane which estimates a delivery predictability metric indicating how likely an entity is to deliver a data to a known destination~\cite{lindgren2003probabilistic}. When two entities meet, they exchange a vector containing the delivery predictabilities for the destinations known to them. The entities increase their metric for each other if they meet often. They also increase the metric for the destinations often encountered by the other entity. Conversely, the metric is decreased if entities do not encounter each other in a while, by the time elapsed since the last encounter. The decision to pass the data when two entities meet is taken if the delivery predictability is higher for the destination of the data at the other entity. The first entity keeps a copy of the data if it has enough free buffer space. Otherwise, the data is dropped according to a FIFO-based policy. 

CAR (Context-aware Adaptive Routing) is a routing protocol that computes and predicts context information at each mobile entity as the main metric for passing the data~\cite{musolesi2005adaptive}. Context information refers to the set of attributes used to optimize the data delivery such as an entity's rate of connectivity change. Each entity maintain a routing table that consists of the delivery probabilities to the other entities. The entities exchange their routing tables when they encounter each other and use them in addition to the context information to update their own routing tables. The update consists of a prediction of the future values of the attributes describing the context using Kalman filter theory for its robustness to missing values in the history and a composition of these estimated values using multi-attribute utility theory. The authors further proposed SocialCast, a publish-subscribe system for mobile entities that adapts the key ideas of CAR to the specificities of the system~\cite{costa2008socially}. In particular, the SocialCast protocol consists of three successive phases: (\textit{i}) dissemination of the interest generated by the entities to their direct neighbors, (\textit{ii}) carrying entity selection according to a utility metric computed for each known interest, and (\textit{iii}) message dissemination by the selected entities. The second phase is the most important and relies on the main idea of CAR to predict the delivery probability using the utilities received from other entities using Kalman filter theory. 

In the same vein, RAPID (Resource Allocation Protocol for Intentional DTN) is a routing protocol that considers routing as a resource allocation problem~\cite{balasubramanian2010replication}. This protocol passes data given its estimated utility with regard to a routing metric such as the worst-case delivery latency, the average delay, or the percentage of packets delivered within a deadline. Since the first entity keeps a copy of the data, passing the data results in replicating the data and thus, degrading performance when resources are limited. The objective of RAPID is to minimize the number of replicas while optimizing a routing metric. The computation of the utility of replication relies on a shared control plane where the entities exchange network state information (\eg expected meeting times with entities and past encounters). In addition to the control information, the entities exchange acknowledgments to remove stale data from their buffer and free up some space to avoid buffer overflows.

MaxProp is a routing protocol for delay tolerant networks designed in the context of the UMass DieselNet~\cite{burgess2006maxprop},  a vehicular network testbed of 40 buses equipped with WiFi capabilities serving the surrounding area of UMass Amherst campus. Similar to PRoPHET and RAPID, with MaxProp, the decision to pass data is made by each bus according to a delivery likelihood estimation based on the history information about the past meetings with other buses. 

While these strategies implement a shared control plane among the entities, they require exchanging large amounts of control information for large networks. Additionally, the shared view of the network is not consistent, as the control information takes time to be propagated in the network.

\subsubsection{\textbf{Towards a centralized control plane}}
\label{sec:centralized-control-plane} Several approaches require a global control plane in order for the nodes to make informed forwarding decisions when in contact. While most approaches implement this control plane in a distributed and \textit{in-band} manner, maintaining a coherent global view of the system at every node is a difficult task. Indeed, this view must describe the state of the system coherently and comprehensively across all the nodes. Some approaches have considered using a centralized control plane to provide such a view. This is the case of BUBBLE, which advocates for centralized community detection that are more efficient than distributed ones, which rely on a partial view of the system~\cite{hui2011bubble}. As noted by the authors of RAPID, using an out-of-band channel to propagate the control plane, enhances the performance in terms of average delivery delay and delivery ratio~\cite{balasubramanian2007dtn}. The out-of-band channel benefits from the instantaneous view of the system state and the bandwidth spared from exchanging the control information between the nodes when in contact. As an example, Polat~\etal leverage a centralized control plane to determine a set of mobile entities that can act as message ferries by their mobility patterns~\cite{polat2011message}. The authors propose heuristics implemented in the control plane to find a maximum-size dominating set of mobile entities connected together. They showed that the entity mobility patterns have a large influence on the performance of the heuristics.


The addition of a control plane helps to increase the performance of the data transfer resulting from the composition of the trajectories of the entities. The control plane gives information on how likely an entity is likely to bring the data closer to its destination. However, in real life, the composition can happen anywhere, but only the contacts that happen in a few specific areas are worth considering to improve the performance of the transfers~\cite{kang2004extracting,sarafijanovic2006island}. The strategies we review in the following section restrict the compositions of the trajectories to pre-positioned locations.

\subsection{Pre-positioned compositions}
\label{sec:indirect-sync-anchored}

\begin{figure}[h]
    \centering
    \includegraphics[width=0.85\columnwidth]{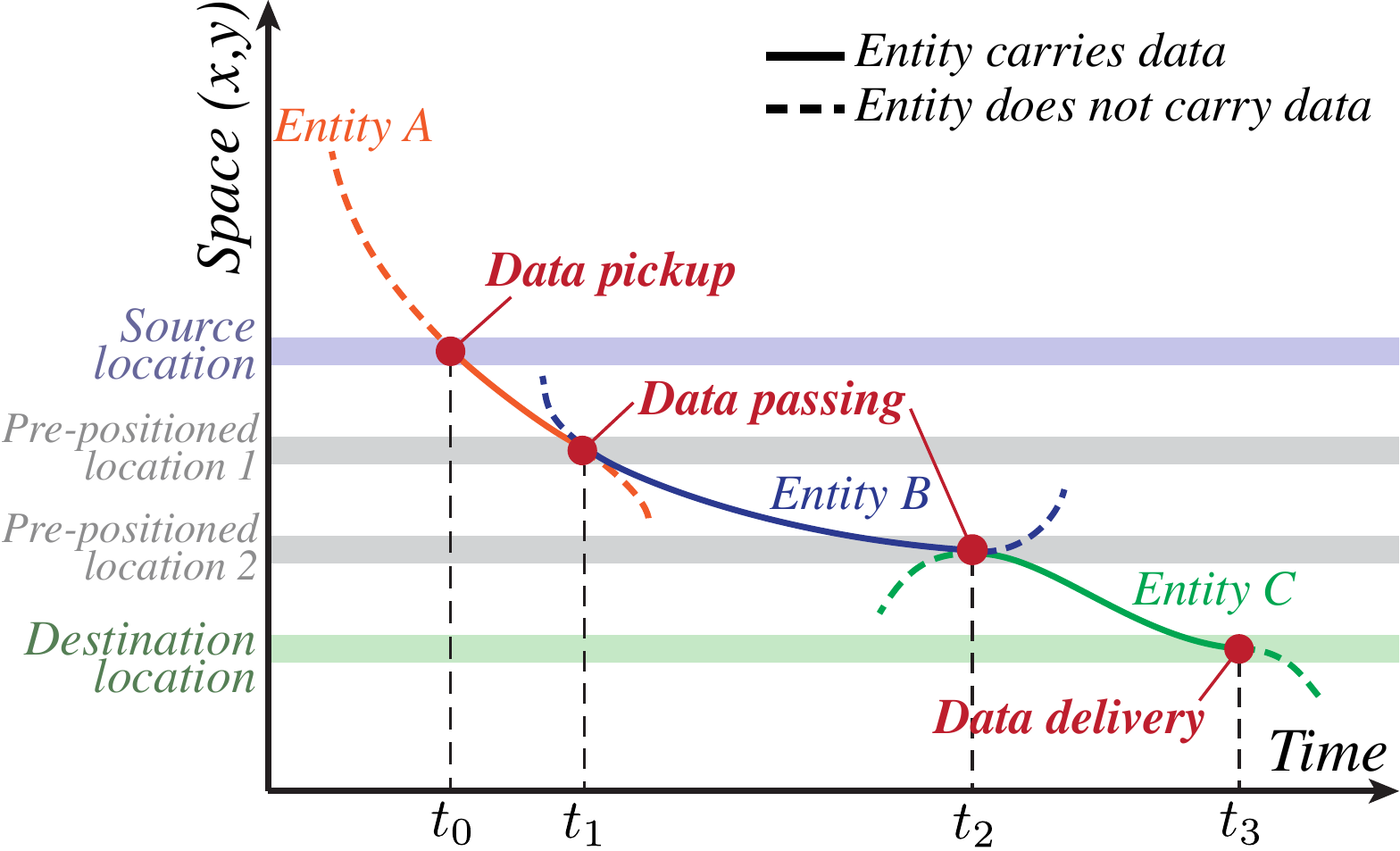}
    \caption{Indirect synchronous composition between entities $A$ and $B$ at pre-positioned location~1, and between entities $B$ and $C$ at pre-positioned location~2.}
    \label{fig:dtn-indirect-sync-geo}
\end{figure}
Instead of passing the data whenever mobile entities meet each other, 
the strategies we review in this section consist in passing the data when the entities meet at pre-positioned locations, 
as shown in Figure~\ref{fig:dtn-indirect-sync-geo}. In this case, the data can be passed between entities only when they are
at passing locations 1 and 2. These strategies exploit the mobility structure of the entities which tend to meet more often 
in these areas. In the case of vehicles, these locations may be road intersections since contacts between vehicles are more likely to happen. 
The identification of these specific areas allows a finer control of the routing strategies over the number of replications needed for one copy to reach the destination. 
Most of the approaches we review in this section create a logical representation of the system, consisting in an overlay graph where the nodes correspond to the areas with high densities of entities and the links represent the movements of the entities between these nodes. This overlay graph is used to route the data across the areas along the shortest path.

\smallskip\noindent\textit{Determining the locations of the passing areas.}
An underlying problem related to pre-positioned passing approach is to determine the locations with high densities of entities. In an urban vehicular setting, they correspond to intersections at junctions and traffic lights. 

LOUVRE~\cite{lee2008louvre} builds an overlay network on top of intersections (``landmarks'') where vehicles are in contact with each other. When the vehicles come to the intersections, they forward the data over the overlay links representing the VANET multi-hop path between two adjacent intersections. Similarly, Tan~\etal uses the movements of vehicles traveling a transportation network to create a ``vehicular backbone network''~\cite{tan2014vehicular}. The vehicles perform ``wireless switching'' when they are in contact at intersections or on dual-way roads between vehicles traveling in opposite directions. 

Sarafijanovic-Djukic \etal characterize the locations with high node densities as ``concentration points'' (CP)~\cite{sarafijanovic2006island}. Using real-life traces, the authors defined CPs as areas visited by at least 5\% of the total vehicles per day. Yuan \etal propose PER, a strategy to forward the data between landmarks which refer to locations where two entities can communicate directly~\cite{yuan2009predict}. However, the authors presuppose the use of pre-positioned landmarks without providing any information on how to determine the landmark locations. 

In the context of multiple message ferries following different routes and creating a communication medium among stationary nodes, Zhao \etal propose to pass the data between the ferries at intersecting points of their routes~\cite{zhao2005controlling}. The routes followed by the ferries must be synchronized among the ferries so as to come in contact regularly. 

Ker{\"a}nen and Ott propose to transport data between airports using the smartphones of the airline passengers~\cite{keranen2009dtn}. The authors rely on the scheduled flight connections at airports to compose the trajectories of the passengers. 

\smallskip\noindent\textit{Maintaining a global view of the network state.}
To be able to create and maintain the logical representation provided by the overlay network, the entities need a consistent and up-to-date view of the network. This view also helps the entities decide where and how to forward the data. There are two approaches to achieve such view. 

The first approach relies on an out-of-band, low-bandwidth control channel to exchange the control messages that are assumed to have a low overhead. This is the case of Tan \etal~\cite{tan2014vehicular} that use such a control channel to have the latest estimation of the overlay link metrics. In the ferry work~\cite{zhao2005controlling}, the ferries use a long range radio to broadcast their route so as they can compute intersecting points where they can meet in an offline manner. 

The second approach leverages an in-band distributed approach. In~\cite{sarafijanovic2006island}, the entities distribute the information about the overlay using a collaborative graph discovery protocol. While this method does not provide an up-to-date information, it avoids relying on signals from the environment. LOUVRE also relies on a peer-to-peer density discovery scheme to populate link state tables at the overlay nodes with the density information~\cite{lee2008louvre}. Similarly, with PER, the entities rely on a partial view of the system by exchanging their history mobility record whenever they meet~\cite{yuan2009predict}. However, in the case of a large-scale network such as the airline network proposed by Ker{\"a}nen and Ott~\cite{keranen2009dtn}, the entities cannot maintain information about all the links in the overlay. The authors argue that epidemic protocols are more suited for this case to forward the data.

\smallskip\noindent\textit{Forwarding data when in the specific areas.}
With the information about the overlay, the entities can make forwarding decisions so the data can follow the shortest path to its destination. With no information on the future trajectories of the entities, the authors of \cite{sarafijanovic2006island} propose to replicate the data to multiple entities to increase the likelihood that at least one will arrive at the next overlay node. Once the data has reached the next node, an acknowledgment is broadcast to the previous node to discard the previous copies of the data. 

Conversely, PER forwards a single copy of the data between the landmarks by using a landmark trajectory prediction to determine the probabilities of contacts between two entities at a given time~\cite{yuan2009predict}. Tan \etal~\cite{tan2014vehicular} forward the data to minimize the delivery delay of the data and guarantee fairness among the data transfers. When two vehicles encounter, they choose the least used overlay links. In LOUVRE, the routing protocol uses the Dijkstra algorithm to forward the data on the overlay links with the highest density of entities.


%% file: 4-discussions.tex
\section{The case of vehicular data offloading}
\label{sec:benj-baron}

In this section, we share our experience with the case of a large-scale offloading system. The system exploits the existing vehicular mobility to transport data and offload large amounts of data from conventional data networks such as the Internet~\cite{baron2017centrally,baron2016offloading}.
While most previous work focused on the forwarding strategies or the prediction of node mobility to enhance the delivery success in sparse networks, our work exploits the large number and wide coverage of the trips made by private vehicles to extend the capacity of conventional data networks, while avoiding costly infrastructure upgrades. With 42 billion vehicle trips made in France yearly, a back-of-the-envelope calculation shows that 10\% of the vehicles traveling the roads of France equipped with a 1~TB hard drive can transport up to 115~EB per day (1.3~PB per second). Extending this idea to the 1.2~billion cars available worldwide, the everyday mobility of vehicles represents an untapped potential for addressing the oncoming data avalanche~\cite{hecht2016bandwidth,index2016forecast}.

Our vehicular offloading system targets transfers of bulk delay-tolerant data, such as background transfers between remote data centers (the communication end-points). The system implements a centralized control plane similar to the works we detailed in Section~\ref{sec:centralized-control-plane}. It consists of three key components: (\textit{i}) a collection of offloading spots that act as stationary intermediate nodes, (\textit{ii}) the offloading overlay, a logical abstraction of the vehicular flows between the offloading spots, and (\textit{iii}) a centralized control plane that dynamically allocates the vehicular resources for the data transfers.

\begin{figure}
    \centering
    \includegraphics[width=\columnwidth]{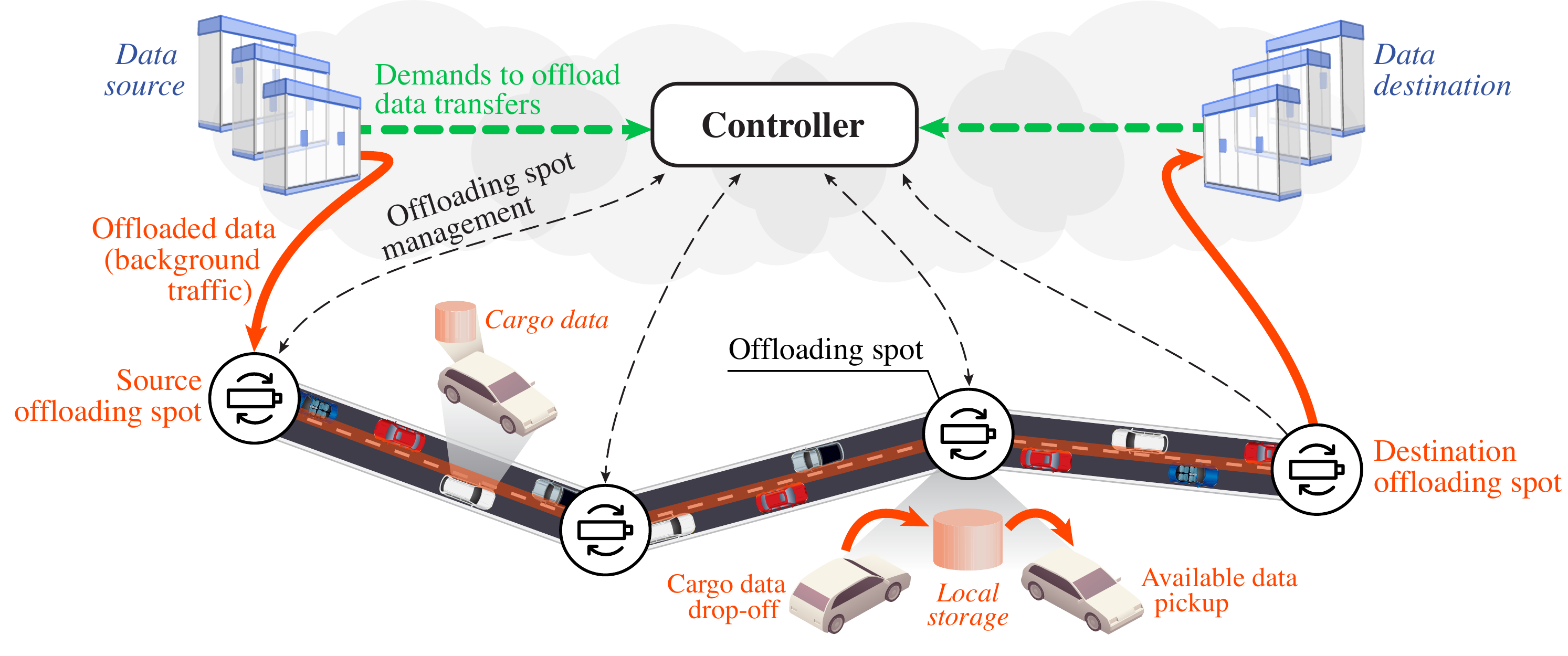}
    \caption{Offloading system operations on a portion of road network.}
    \label{fig:offloading-system}
\end{figure}

\smallskip\noindent\textit{Asynchronous pre-positioned compositions at stationary offloading spots.} The data offloading system relies on a collection of offloading spots deployed at charging stations in the case of electric vehicles. Offloading spots act as data exchange relay points where vehicles drop off their data cargo for later pick-ups. The vehicle transfers the data between its on-board storage and the offloading spot's storage with state-of-the-art high-throughput wireless technologies (\eg MIMO 802.11ac) while the vehicle is charging its battery. When reaching the last offloading spot (the closest to the destination), the data needs to be transloaded to the destination. We consider that the delivery is successful when the vehicle reaches the last offloading spot and the hard drive is taken out of the vehicle. The data follows a vector path consisting of a sequence of offloading spots and the flow of vehicles connecting them together. The offloading spots enable the asynchronous composition of the vehicles' trajectories. We represent the main operations of the system in Figure~\ref{fig:offloading-system}.

\smallskip\noindent\textit{Characterizing the entity movements into network quantities.} 
Flows of vehicles carry the data between offloading spots toward the destination of the data. A logical view of the offloading spots and the flows of vehicles connecting them allows to treat the flows as a networking resource to achieve data transfers. This view (\textit{i}) mitigates the complexity of the road network and (\textit{ii}) translates the vehicle movements into networking quantities. As a result, the view enables the efficient allocation of data transfers to the flows of vehicles. 

\smallskip\noindent\textit{Centralized control plane.} Instead of using a distributed control plane as it is the case with most of the related approaches, the system leverages a centralized architecture with a controller in the vein of the software-defined networking (SDN) paradigm~\cite{mckeown2008openflow}. The controller relies on an out-of-band communication channel connected to the offloading spots. Thus, the controller has a holistic view of the offloading infrastructure to efficiently allocate the vehicular resources for the data transfers. In particular, the allocation aims to maximize the performance requirements of the transfers while guaranteeing a fair distribution of the road resources to the data transfers. The controller then derives forwarding states from the allocation of the data transfers and installs them at the offloading spots. With these states installed, the offloading spots are able to decide which locally-available data the stopped vehicles can pickup. To this end, the offloading spot needs to predict the vehicle's future route and destination, which can be done using historical data and the vehicle's navigation system with the driver's permission. The vehicle will then pickup the data whose destination matches the vehicle's predicted route. Most of the data delivery approaches rely on redundant data copies to increase the delivery ratio. However, since redundancy does not guarantee reliable data transfers, the centralized approach enables reactive error control by retransmitting the data that did not reach the expected destination.

The logical view, together with the centralized architecture, enables the efficient allocation and management of the vehicular resources for data transfers offloaded on the road network. With an instance of the system on the roads of France, they showed that the road network has the potential to offload several Petabytes of data per day.

Our approach is promising as it solves the  problem of most of the approaches that relied on a distributed architecture to compute the vector route such as RAPID~\cite{balasubramanian2007dtn}. With its centralized control, the approach is the first that leverages an out-of-band channel to dynamically modify the behavior of the offloading spots in order to adapt the vehicular resources and current load to the data transfer demands in a timely manner.

\input{table_classification_metrics_r2.tex}

\section{Metrics and evaluation tools}
\label{sec:metrics-eval-tools}

The range of applications and entities involved in the data transfers makes it difficult to compare the different approaches. The metrics needed to assess the performance of the data transfers vary with the applications and the tools and mobility models vary with the entities under consideration. In this section, we give an overview of the metrics and mobility models we believe are important for the evaluation of the different approaches. We summarize in Table~\ref{tab:metrics} the works we reviewed in the survey according to their data delivery approaches, as well as the performance metrics and evaluation tools they consider.

\subsection{Performance metrics}
\label{sec:perf-metrics}

In the following, we present the metrics used by the different works to assess the performance of the data transfers relying on the strategies they propose. These metrics are also used to compare the different strategies with the proposed one under the same experimental scenario. In Table~\ref{tab:classification-metrics}, we give a summary of the performance of the data delivery approaches we reviewed in the survey according to the most relevant metrics listed below. This table was inspired from the one presented by Shah~\etal~\cite{shah2003data}. Note that we present rough estimations of the performance metrics that correspond to the typical results we found in the literature we have surveyed.

With their high delivery ratio, high throughput, and low delivery delay, the traditional data delivery approaches relying on wired and cellular infrastructures are well suited for real-time communications and large data transfers between regions with high-throughput connectivity. However, these approaches are limited by their capacity to handle current traffic demands~\cite{hecht2016bandwidth} and their coverage in remote regions. Indeed, they require deploying expensive infrastructure, thus limiting their coverage and operation in areas with low densities of populations (\eg rural areas) and in remote places (\eg battlefields, disaster relief, and wide-area sensing). These latter scenarios benefit from the use of the alternative data delivery approaches we have surveyed. Compared to traditional systems, these alternative approaches \textit{trade} performance for enabling communications where they could not have been possible.

\input{table_metrics_r2.tex}

\smallskip\noindent\textit{Delivery delay.}
A data transfer is generally characterized in terms of \textit{delivery delay}, that is, the average delay to deliver the data from the sources to the destinations. For instance, the epidemic routing strategy yields the lower bound of the delivery delay, as the data is replicated when passed and one copy will follow the shortest path between the source and the destination~\cite{vahdat2000epidemic}. 

\smallskip\noindent\textit{Delivery ratio (or rate).} The delivery ratio of a data transfer corresponds to the proportion of data successfully delivered to the destination compared to the amount of data generated at the source. With limited storage capacities, the entities rely on buffer management strategies to select the data to drop when composing the trajectories~\cite{balasubramanian2007dtn,juang2002energy}, which leads to data losses. Additionally, with deadlines or timeouts, the data can expire while being carried by an entity~\cite{hui2011bubble}. In this case, the data is dropped and considered as lost. To improve the reliability of the data transfers, the data passing strategies aim to minimize the data losses measured at the application level. 

\smallskip\noindent\textit{Number of copies (protocol overhead).}
When using replication to compose trajectories, multiple copies of the data follow different logical paths to eventually reach the destination~\cite{vahdat2000epidemic,spyropoulos2005spray}. While replicating reduces the delivery delay, it requires more storage resources at the entities to store the additional copies, as well as more bandwidth during contacts to forward them. Transmitting additional data copies creates an overhead that the different data delivery approaches aim to minimize.

\smallskip\noindent\textit{Hop count.} 
In the case of indirect data delivery, the average number of hops traveled by the data corresponds to the average number of entities that successively carried the data from its source to its destination. The strategies that compose the trajectories aim to minimize the hop count and lower the delivery delay. To this end, the composition is done so as to find the entity that will carry the data to its destination with the fewest number of hops.

\smallskip\noindent\textit{Energy efficiency.} 
The storage and the communication interfaces that equip the entities are generally powered by batteries with limited capacity. The energy efficiency is measured as the average amount of data delivered per unit energy consumption. Works studied the energy impact of composing the trajectories in the lifetime of the entities~\cite{zhao2004message,juang2002energy}. Since high-bitrate radios consume more energy than storage, the energy-aware strategies that compose the trajectories aim to minimize the number of hops and copies generated for a data. 

\smallskip\noindent\textit{Throughput and data traffic load.}
A data delivery system is also characterized by the throughput it can achieve. The throughput refers to the amount of data that can be successfully transferred between a source and a destination per unit of time. The throughput of a system is usually measured indirectly in combination with another performance metric (\eg delivery delay or delivery ratio) under varying loads of data to transfer~\cite{balasubramanian2007dtn}. With constrained resources, varying the traffic load allows stressing the system by filling the buffers of the entities and the bandwidth of the wireless transmissions. In the indirect data delivery case, the different approaches must make efficient utilization of the composition of the entity trajectories to increase the system throughput~\cite{grossglauser2001mobility,chen2015dtn}.  Additionally, this metric helps compare the performance of data transfers relying on entity movements with those achieved by infrastructure-based networks such as the Internet~\cite{zarafshan2010trainnet,tan2014vehicular}. 

\smallskip\noindent\textit{Fairness.}
Fairness issues generally arise when multiple data flows share the resources created by the movements of the entities. A data flow is characterized by its source and destination. In the case of direct data delivery, the entity must visit all the data sources and destinations to avoid any starving data flows. Conversely, in the case of indirect data delivery, the data passing strategies must also guarantee the same chances to transfer data for the different flows. Fairness is generally measured against a performance metric $X$ using Jain's fairness index given by~\cite{jain1984quantitative}:
\[
    \frac{X_{i}^{2}}{n\big(\sum_{i}X_{i}\big)^{2}}\comma
\]
\noindent where $X_i$ is the value of the performance metric $X$ for data flow $i$ or a copy of the data (\eg throughput of a data flow or delivery delay of the copies) and $n$ is the total number of flows. The result ranges from $\nicefrac{1}{n}$ (worst case) to 1 (best case).

\smallskip\noindent\textit{Deployment cost.} 
Equipping the entities with storage and communication capabilities has a non-negligible cost. Some work detailed the costs of equipping the entities in the context of real deployments~\cite{seth2006low,pentland2004daknet,burgess2006maxprop}. Deploying a dedicated infrastructure to support the composition of the trajectories has also a non-negligible cost; however, none of the works propose a deployment so as to minimize its cost. Moreover, some works propose to rely on existing delivery services, whose costs result from the use of the services~\cite{cho2011budget,laoutaris2013delay}. In this case, the data delivery strategies aim to minimize the cost of the delivery. 

\subsection{Reproducing the mobility of entities}
\label{sec:mobility-models}

Most of the approaches rely on mobility models to assess the performance of the data delivery strategies. As stated by Juang \etal~\cite{juang2002energy}: ``Mobility models help to abstract how fast and how often the [entities] move, in what direction, and with what forces of attraction or repulsion.'' One needs to understand how the entities will move, as this critically affects hardware, protocol, and overall system design. In the following, we review the different mobility models used to assess the performance of the different data delivery strategies.

Synthetic mobility models can be extracted from mathematical stochastic processes, such as Random WayPoint~\cite{johnson1996dynamic} and Random Walk~\cite{pearson1905problem}. However, while these models are simple to understand and analyze, they are not realistic for real-life situations. With these models, the entities have (\textit{i}) the same mobility characteristics and (\textit{ii}) they can move equally frequently to every location within the area of interest~\cite{sarafijanovic2006island}. 

Several works have extensively studied mobility traces from real-world situations such as university campuses, conferences, or taxi movements in urban environments. With these works, the authors were able to extract and generalize characteristics of the entity movements. They showed that the two assumptions that characterize the simple mobility models do not hold true. Specifically, these studies showed properties at different levels of granularity. At the entity level, they showed that the entities have a non-homogeneous spatial distribution and that the entities tend to visit some places more than others~\cite{mcnett2005access}. At the pair-wise level, they showed that two entities have a skewed, power law inter-meeting time distribution~\cite{chaintreau2007impact}. Finally, at the community level, they showed that the entities tend to gather together to form communities, with some entities more popular than others (\ie with a high centrality)~\cite{juang2002energy,hui2011bubble}. 

These studies led to define more realistic mobility models to characterize the entity behavior. With these models, the authors of the different works were able to assess the performance of the data delivery approaches under realistic scenarios with a heterogeneous spatial distribution of entities. In parallel, the works also improved the composition of the trajectories to fit with the realistic mobility models by taking into account the properties of the entity movements and encounters.


\subsection{Evaluation tools}

Simulations and deployments are the main way that authors use to assess and compare the performance of the data delivery strategies they propose. As we can see in Table~\ref{tab:metrics}, multiple discrete-event simulators are available to simulate the mobility of the entities and the packet-level interaction of the data delivery protocols. ONE is used to simulate the mobility of the entities with models already implemented~\cite{keranen2009one}. It also allows easy implementation of data delivery protocols by overriding primitives when the entities come in contact or finish one. However, because of its simplicity, ONE does not implement the mechanisms usually found at link-layer, such as wireless interferences due to buildings and other communications taking place in the vicinity, or bandwidth variation due to the distance between the entities. \textit{ns-2} and its evolution, \textit{ns-3}, are also very popular simulators mainly used for packet-level analysis~\cite{mccanne1997network,riley2010ns}. While they both simulate realistic link-layer wireless protocols, they do not take into account an extensive range of mobility models for the entities. SUMO and OMNET++ are also used to model the mobility of vehicles traveling the roads~\cite{SUMO2012,varga2001omnet}. SUMO provides a realistic simulation of the vehicles and their interactions on a road network and OMNET++ implements a network-layer to enable the communication between the vehicles. 

However, these simulators are heavy, as they require large memory and processing power. They turn out to be inefficient when it comes to simulate entity movements in a large-scale dataset. As a result, a large body of works rely on custom simulators that were designed specifically for the protocol(s) and the mobility model(s) under evaluation. With multiple custom simulators used and their variable availability to the open-source community, this creates a problem to reproduce the experiments and their results. Note that the limited availability of some mobility datasets also hinders the reproducibility of the results.

%% file: table_classification_metrics_r2.tex
\begin{table*}[t]
  \small
  \centering
    	\caption{Various performance metrics for the different data delivery approaches in the classification.}
  \resizebox{2\columnwidth}{!}{%
  \begin{tabular}{p{0.5cm}L{3.5cm}L{2.2cm}L{2cm}L{2cm}L{2cm}L{2cm}L{2.5cm}L{1.5cm}} 
  \toprule  
  
  \multicolumn{2}{l}{\textbf{Data delivery approaches}} & \textbf{Data delivery delay} & \textbf{Data delivery ratio} & \textbf{Resulting throughput} & \textbf{Energy consumption} & \textbf{Buffer memory} & \textbf{Deployment cost} & \textbf{Distance covered}\\
  \midrule  
  
  \multicolumn{2}{l}{Infrastructure-based (Internet)} & \makecell[l]{Low \\(100-300 ms)} & High (99.9\%) & High (Gbps) & High & Low & High & Global \\[3pt]
  
  \midrule  

  \multicolumn{2}{l}{Base station (cellular)} & \makecell[l]{Low \\(100-1000 ms)} & High (99\%) & \makecell[l]{High\\ (100 Mbps)} & High & Low & High & National \\[3pt]
  
  \midrule  
  
  \multicolumn{2}{l}{Ad-hoc (MANET)} & Medium (seconds) & Medium (50-80\%) & \makecell[l]{Medium\\ (1 Mbps)} & Medium & Medium & Medium-High & Local \\[3pt]
    
  \midrule  

  \multirow{3}{*}{\rotatebox[origin=c]{90}{\parbox[c]{2cm}{\centering Direct data delivery}}} & Passive & High (hours) & Low (20-50\%) & Low (kbps) & Low & High & Low & Local \\
  
  \cmidrule(lr){2-9}

  & Active & Medium (minutes-hours) & High (90\%) & Low (kbps) & High & High & High (mobile relay costs) & Local \\
  
  \cmidrule(lr){2-9}
  
  & Paying  & High (days) & High (90\%) & High (Gbps) & High & High & Medium (postal service fees) & Global \\
  
  \midrule  
  
  \multirow{4}{*}{\rotatebox[origin=c]{90}{\parbox[c]{3cm}{\centering Indirect data\\delivery}}} & Asynchronous data passing (stationary relay nodes) & Medium (minutes-hours) & Medium (50-80\%) & High (Gbps) & Medium-High & High & Medium (relay node costs) & National \\
  
  \cmidrule(lr){2-9}
  
  & Asynchronous data passing (mobile relay nodes) & Medium (minutes) & Medium (50-80\%) & Low (kbps) & Medium-High & High & High (mobile relay costs) & Local \\
  
  \cmidrule(lr){2-9}
  
  & Synchronous data passing (floating locations) & Medium (minutes) & Medium (50-80\%) & Low (kbps) & Medium & Low-Medium & Low & Local \\
  
  \cmidrule(lr){2-9}

  & Synchronous data passing (pre-positioned locations) & Medium (minutes-hours) & Low (20-50\%) & High (Mbps) & Low & Low-Medium & Low & Local \\
	 
 \bottomrule

  \end{tabular}
  }
  	\label{tab:classification-metrics}
\end{table*}

%% file: table_metrics_r2.tex
\bigskip
\renewcommand{\arraystretch}{1.3}
\setlength{\tabcolsep}{5pt}

\begin{table*}[ht!]
  \small
  \centering
  \caption{Summary of key data delivery approaches.}
  
  \vspace{-5pt}
  \resizebox{2\columnwidth}{!}{%
  \begin{tabular}{c | c | c | c | c} 
	& & & & \\[-7pt]
	
	\toprule
	\multirow{2}{*}{} \textbf{Reference}  & \textbf{Delivery} & \textbf{Data passing} & \textbf{Performance metrics covered} & \textbf{Performance assessment} \\
	\midrule
	\addlinespace[0pt]


	\cite{grossglauser2001mobility} Grossglauer and Tse & \cellcolor[gray]{0.92} Direct & \cellcolor[gray]{0.92} Passive & Per-node throughput & Analytical model, Simulation (unknown) \\
	\cite{jain2004routing} First Contact & \cellcolor[gray]{0.92} Direct & \cellcolor[gray]{0.92} Passive & Delivery ratio, delay & Simulations (Java) \\ 
	\cite{burrell2004vineyard} Burrel \etal  & \cellcolor[gray]{0.92} Direct & \cellcolor[gray]{0.92} Passive & None & Deployment \\
	\cite{mcdonald2007sensor} SeNDT  & \cellcolor[gray]{0.92} Direct & \cellcolor[gray]{0.92} Passive & None & Deployment \\
	\cite{lahde2007practical} EMMA project & \cellcolor[gray]{0.92} Direct &
	\cellcolor[gray]{0.92} Passive & Throughput, success ratio & Deployment \\
	\cite{pentland2004daknet} DakNet & \cellcolor[gray]{0.92} Direct & 
	\cellcolor[gray]{0.92} Passive & Cost, per-node throughput & Deployment \\
	\cite{rabagliati2004wizzy} Wizzy Digital Courier & \cellcolor[gray]{0.92} Direct &
	\cellcolor[gray]{0.92} Passive & Cost, per-node throughput & Deployment\\
	\cite{rfc1149} RFC IPoAC & \cellcolor[gray]{0.92} Direct & 
	\cellcolor[gray]{0.92} Passive & Delivery ratio, delay & Deployment \\
	
	\cite{zhao2003message} Message ferry  & \cellcolor[gray]{0.98} Direct & \cellcolor[gray]{0.98} Active & Delivery ratio, delay & Simulation (unknown) \\ 
	\cite{mansy2011deficit} Mansy \etal & \cellcolor[gray]{0.98} Direct & \cellcolor[gray]{0.98} Active & Delivery delay & Simulation (ONE) \\ 
	\cite{tirta2004efficient} Tirta \etal & \cellcolor[gray]{0.98} Direct & \cellcolor[gray]{0.98} Active & Delivery delay, energy efficiency & Simulation (NS-2) \\
	\cite{zhao2005controlling} Message ferries & \cellcolor[gray]{0.98} Direct & \cellcolor[gray]{0.98} Active & Delivery ratio, energy efficiency & Simulation (unknown) \\ 
	\cite{zhou2013minimizing} Pigeon networks  & \cellcolor[gray]{0.98} Direct & \cellcolor[gray]{0.98} Active &  Delivery delay & Analytical model\\ 
	\cite{tekdas2009using} Tekdas \etal & \cellcolor[gray]{0.98} Direct & \cellcolor[gray]{0.98} Active & Delivery ratio, energy efficiency & Deployment  \\
	
	\cite{wang2004turning} Postmanet & \cellcolor[gray]{0.92} Direct & \cellcolor[gray]{0.92} Paying & None & None \\
	\cite{laoutaris2013delay} Laoutaris \etal & \cellcolor[gray]{0.92} Direct & \cellcolor[gray]{0.92} Paying & Throughput, cost & Simulation (unknown) \\
	
	\cite{zhao2006capacity} Throwboxes & \cellcolor[gray]{0.98}Indirect & \cellcolor[gray]{0.98}Async. stationary & Delivery ratio, delay & Simulation (NS-2) \\ 
	\cite{chawathe2006inter} Dead-Drops & \cellcolor[gray]{0.98}Indirect & \cellcolor[gray]{0.98}Async. stationary & Delivery delay & Analytical model \\	
	\cite{ibrahim2009analysis} Ibrahim \etal & \cellcolor[gray]{0.98}Indirect & \cellcolor[gray]{0.98}Async. stationary & Delivery delay, overhead & Simulation (unknown) \\
	\cite{seth2006low} KioskNet & \cellcolor[gray]{0.98}Indirect &\cellcolor[gray]{0.98} Async. stationary & None & Deployment \\
	\cite{demmer2007dtlsr} DTLSR &\cellcolor[gray]{0.98} Indirect & \cellcolor[gray]{0.98}Async. stationary & Delivery ratio, delay & Simulation (C++) \\
	\cite{chen2015dtn} DTN-FLOW & \cellcolor[gray]{0.98} Indirect & \cellcolor[gray]{0.98} Async. stationary & Delivery ratio, delay, overhead & Simulation (unknown) \\
	\cite{zarafshan2010trainnet} TrainNet & \cellcolor[gray]{0.98} Indirect & \cellcolor[gray]{0.98} Async. stationary & Delivery ratio, delay, overhead, fairness & Simulation (DESMO-J) \\
	\cite{banerjee2007energy} Throwboxes & \cellcolor[gray]{0.98} Indirect & \cellcolor[gray]{0.98} Async. stationary & Delivery ratio, delay, energy efficiency & Deployment, Simulation (Java) \\
	\cite{wang2004turning} Postmanet & \cellcolor[gray]{0.98} Indirect & \cellcolor[gray]{0.98} Async. stationary & Delivery delay & Simulation (unknown) \\ 
	\cite{cho2011budget} Cho and Gupta & \cellcolor[gray]{0.98} Indirect & \cellcolor[gray]{0.98} Async. stationary & Cost, delivery delay & Simulation (unknown) \\
	\cite{baron2017centrally} Baron \etal & \cellcolor[gray]{0.98} Indirect & \cellcolor[gray]{0.98} Async. stationary & Per-flow throughput & Simulation (unknown)  \\
	
	\cite{zhao2004message} Message ferry & \cellcolor[gray]{0.92} Indirect & \cellcolor[gray]{0.92} Async. mobile & Delivery ratio, delay, energy efficiency & Simulation (NS) \\ 
	\cite{bin2006message} Bin \etal & \cellcolor[gray]{0.92} Indirect & \cellcolor[gray]{0.92} Async. mobile & Delivery delay, fairness & Simulation (unknown)  \\
	\cite{burns2008mora} Burns \etal  & \cellcolor[gray]{0.92} Indirect & \cellcolor[gray]{0.92} Async. mobile & Delivery ratio, delay & Simulation (unknown) \\ 
	\cite{asadpour2016route} Asadpour \etal & \cellcolor[gray]{0.92} Indirect & \cellcolor[gray]{0.92} Async. mobile & Delivery ratio, delay, hops & Deployment \\
	
	\cite{vahdat2000epidemic} Epidemic routing & \cellcolor[gray]{0.98} Indirect & \cellcolor[gray]{0.98} Sync. floating & Delivery ratio, delay & Simulation (NS-2)\\ 
    \cite{papadopouli2001design,papdopouli2000seven} 7DS & \cellcolor[gray]{0.98} Indirect & \cellcolor[gray]{0.98} Sync. floating & Delivery ratio & Simulation (NS-2)\\
	\cite{shah2003data} Data MULEs & \cellcolor[gray]{0.98} Indirect & \cellcolor[gray]{0.98} Sync. floating & Delivery ratio & Analytical model, Simulation (unknown)\\ 
	\cite{hull2006cartel} CarTel & \cellcolor[gray]{0.98} Indirect & \cellcolor[gray]{0.98} Sync. floating & Throughput & Deployment\\
	\cite{eisenman2009bikenet} BikeNet & \cellcolor[gray]{0.98} Indirect & \cellcolor[gray]{0.98} Sync. floating & Throughput, delivery ratio, delay & Deployment\\ 
	\cite{goodman1997infostations} Infostations & \cellcolor[gray]{0.98}Indirect & \cellcolor[gray]{0.98} Sync. floating & Throughput, energy efficiency & Analytical model\\
	\cite{small2003shared} SWIM & \cellcolor[gray]{0.98}Indirect & \cellcolor[gray]{0.98} Sync. floating & Delivery ratio, delay & Simulation (unknown)\\ 
	\cite{leontiadis2007geopps} GeOpps & \cellcolor[gray]{0.98}Indirect & \cellcolor[gray]{0.98} Sync. floating & Delivery ratio, hops, delay, overhead & Simulation (OMNET++)\\
	\cite{soares2014geospray} GeoSpray & \cellcolor[gray]{0.98}Indirect & \cellcolor[gray]{0.98} Sync. floating & Delivery ratio, hops, delay, overhead &  Simulation (ONE)\\
	\cite{jain2004routing} Jain \etal & \cellcolor[gray]{0.98}Indirect & \cellcolor[gray]{0.98} Sync. floating & Delivery ratio, delay & Simulations (Java)\\
	\cite{spyropoulos2005spray} Spray-and-Wait & \cellcolor[gray]{0.98}Indirect & \cellcolor[gray]{0.98} Sync. floating & Delivery ratio, delay, overhead & Simulation (unknown)\\
	\cite{spyropoulos2007spray} Spray-and-focus & \cellcolor[gray]{0.98}Indirect & \cellcolor[gray]{0.98} Sync. floating & Delivery delay, overhead & Simulation (unknown)\\
	\cite{davis2001wearable} DLE & \cellcolor[gray]{0.98}Indirect & \cellcolor[gray]{0.98} Sync. floating & Delivery ratio & Simulation (unknown)\\	
	\cite{juang2002energy} ZebraNet & \cellcolor[gray]{0.98}Indirect & \cellcolor[gray]{0.98} Sync. floating & Delivery ratio, energy efficiency & Simulation (C)\\ 
	\cite{hui2011bubble} BUBBLE & \cellcolor[gray]{0.98}Indirect & \cellcolor[gray]{0.98} Sync. floating & Delivery ratio, overhead & Simulation (OMNET++)\\
    \cite{musolesi2005adaptive} CAR & \cellcolor[gray]{0.98}Indirect & \cellcolor[gray]{0.98} Sync. floating & Delivery ratio, delay, overhead & Simulation (OMNET++)\\
    \cite{costa2008socially} SocialCast & \cellcolor[gray]{0.98}Indirect & \cellcolor[gray]{0.98} Sync. floating & Delivery ratio, delay, overhead & Simulation (OMNET++)\\
	\cite{lindgren2003probabilistic} PRoPHET & \cellcolor[gray]{0.98}Indirect & \cellcolor[gray]{0.98} Sync. floating & Delivery delay, overhead & Simulation (unknown)\\
	\cite{balasubramanian2010replication} RAPID & \cellcolor[gray]{0.98}Indirect & \cellcolor[gray]{0.98} Sync. floating & Delivery ratio, delay, overhead, fairness &  Simulation (unknown)\\ 
	\cite{burgess2006maxprop} MaxProp & \cellcolor[gray]{0.98}Indirect & \cellcolor[gray]{0.98} Sync. floating & Delivery ratio, delay & Deployment, Simulation (unknown)\\ 
	
	\cite{lee2008louvre} LOUVRE & \cellcolor[gray]{0.92} Indirect & \cellcolor[gray]{0.92} Sync. pre-positioned & Delivery ratio, delay, overhead & Simulation (Qualnet) \\
	\cite{tan2014vehicular} Tan \etal & \cellcolor[gray]{0.92}Indirect & \cellcolor[gray]{0.92} Sync. pre-positioned & Throughput, delivery delay & Simulation (Matlab)\\
	\cite{sarafijanovic2006island} Island Hopping & \cellcolor[gray]{0.92}Indirect & \cellcolor[gray]{0.92} Sync. pre-positioned & Delivery ratio, delay, overhead & Simulation (unknown) \\
	\cite{yuan2009predict} PER & \cellcolor[gray]{0.92}Indirect & \cellcolor[gray]{0.92} Sync. pre-positioned & Delivery ratio, delay & Simulation (Java) \\
	\cite{zhao2005controlling} Message ferries  & \cellcolor[gray]{0.92}Indirect & \cellcolor[gray]{0.92} Sync. pre-positioned & Delivery ratio, energy efficiency & Simulation (unknown) \\
	\cite{keranen2009dtn} Ker{\"a}nen and Ott  & \cellcolor[gray]{0.92}Indirect & \cellcolor[gray]{0.92} Sync. pre-positioned & Delivery ratio, delay, hops & Simulation \\ 

	\addlinespace[-2pt]
 	\bottomrule

  \end{tabular}
  }
  	\label{tab:metrics}
    \vspace*{-20pt}
\end{table*}

%% file: 5-conclusion.tex
\section{Conclusion and open research challenges}
\label{sec:conclusion}

In this paper, we surveyed the literature on data transfers relying on the mobility of everyday entities to  physically transfer data. We proposed a classification of the works based on whether the data delivery is direct and achieved by a single entity, or it is indirect and achieved using the movements of several entities. In the case of indirect data delivery, we further categorized the works depending on the time and space criteria used for the composition of the entity trajectories. We additionally developed a generic methodology to characterize and use the mobility of the entities to provide value-added services. Our survey provides a comprehensive overview of the state of the art in the use of mobility to transport data and allows outlining the following open research directions.

\textit{Real-world traces}. With more comprehensive datasets available, we see shifts in the mobility models of entities, from synthetic to more recent data-driven models. This enables more complex dynamic models of entity mobility that evolve with time. The entity movements become more predictable, allowing the approaches to use the known or predictable movements of some entities to mimic the deterministic behaviors of intermediate controllable nodes.

\textit{Centralized control}. With the availability of low-power and long range access networks such as SigFox\footnote{SigFox t \url{http://www.sigfox.com/}} or LoRa\footnote{LoRa Alliance at \url{https://www.lora-alliance.org/}}), an increasing amount of works propose to rely on an out-of-band control channel to orchestrate and coordinate the entities. Indeed, the global view of the system cannot be achieved in an in-band manner for obvious lack of consistency. A central controller is then in charge of gathering the current state of the system, using it to feed predictive mobility models, and configuring the entities to behave accordingly.

\textit{Management issues}. Management of the alternative communication channel refers to both the configuration of the control plane of the entities and their monitoring. While the works we surveyed omit this aspect, management is an important task of deploying and running a network. The management issues are facilitated with a centralized control plane, however, they arise with uncontrolled entities.

\textit{Comparison of various entities}. Each work we surveyed focused on one type of entity dedicated to provide the communication for a set of services. However, no work has compared the performance of different entities for the same target service. In this case, we need different performance metrics to assess whether one entity is better than the other, \eg reach (local citywide or global countrywide), amount of data that can be transported by one entity, delivery delay (depends on the reach), delivery rate, or throughput.

\textit{Multi-modal data transport}. Combining the mobility of various entities at different scales would create a multi-tiered architecture involving intermodal end-to-end data transfers, ranging from large-scale (\eg country-wide) to local (\eg within a city) end-to-end data transfers.

\textit{Data and computation offloading}. The mobility and the resources available at the entities could benefit services by enabling better distribution of the compute resources and reduce latency. Instead of using cellular access networks and cloud computing resources, services could leverage the resources available in the vicinity of the mobile nodes.

\textit{Data transfer security and privacy.} Since the data is physically transferred over an alternative channel that consists of mobile entities, the only way have access to the traffic would be to hijack the entities on their way to the destinations. A way to prevent the hijacker from accessing the data is to encrypt the content transferred. As so, with less points of failure, the resulting data transfers would be less subject to security and privacy breaches as compared to legacy data transfers on infrastructure-based networks.

We believe that these challenges outline key future research directions that leverage mobile entities as alternative means of data delivery.